\documentclass[prx,twocolumn,superscriptaddress,showpacs,notitlepage]{revtex4-1}
\usepackage{amsmath,amssymb}
\usepackage{graphicx}
\usepackage{txfonts}
\usepackage{color}
\usepackage{hyperref}
\usepackage{framed} 
\usepackage{wrapfig}
\usepackage[normalem]{ulem}
\usepackage[a4paper]{geometry}
\newcommand{\edit}[1]{{\color{blue}{#1}}}    

 \definecolor{ashgrey}{rgb}{0.7, 0.75, 0.71}
 
\global\long\def\ket#1{\left| #1\right\rangle }
 \global\long\def\bra#1{\left\langle #1 \right|}
 \global\long\def\av#1{\left\langle #1 \right\rangle }

\begin{document}

\title{ Transient phases and dynamical transitions in the post quench evolution of the generalized Bose-Anderson model}

\author{Dmitry V. Chichinadze}
\email[Correspondence should be addressed to ]{dchichinadze@gmail.com}
\affiliation{Department of Physics, Lomonosov Moscow State University, Leninskie gory 1, 119991 Moscow, Russia}
\affiliation{Russian Quantum Center, Novaya 100, 143025 Skolkovo, Moscow Region, Russia}
\author{Pedro Ribeiro}
\affiliation{CeFEMA, Instituto Superior T\'ecnico, Universidade de Lisboa Av. Rovisco Pais, 1049-001 Lisboa, Portugal}
\affiliation{Russian Quantum Center, Novaya 100, 143025 Skolkovo, Moscow Region, Russia}
\author{Yulia E. Shchadilova}
\affiliation{Department of Physics, Harvard University, Cambridge, Massachusetts 02138, USA}
\affiliation{Russian Quantum Center, Novaya 100, 143025 Skolkovo, Moscow Region, Russia}
\author{Alexey N. Rubtsov}
\affiliation{Russian Quantum Center, Novaya 100, 143025 Skolkovo, Moscow Region, Russia}
\affiliation{Department of Physics, Lomonosov Moscow State University, Leninskie gory 1, 119991 Moscow, Russia}


\begin{abstract}

The exact description of the time evolution of open correlated quantum systems remains one of the major challenges 
of condensed matter theory, specially for asymptotic long times where most numerical methods fail.
Here, the post-quench dynamics of the $N$-component Bose-Anderson impurity model is studied in the $N\to\infty$ limit. 
The equilibrium phase diagram is similar to that of the Bose-Hubbard model in that it contains local versions of  Mott and Bose phases. 
Using a numerically exact procedure we are able to study the real time evolution including asymptotic long time regimes.
The formation of long-lived transient phases is observed for quench paths crossing foreign phases. 
For quenches inside the local Bose condensed phase, a dynamical phase transition is reported, that separates the
evolution towards a new equilibrium state and a regime characterized at large times by a persistent phase rotation of the order parameter.
We explain how such non-decaying mode can exist in the presence of a dissipative bath. 
We discuss the extension of our results to the experimental relevant finite-$N$ case and their implication for the existence of non-decaying modes in generic quantum systems in the presence of a bath. 
\end{abstract}

\maketitle

\section{Introduction}
\label{sec:1}

The study of  out-of-equilibrium processes in correlated quantum systems is a rapidly growing field. 
Progress has been driven by experimental advances in solid state and ultra-cold atomic setups that made possible the preparation, manipulation and probing of quantum many-body states in real time.
On the theory side, the study of correlated systems far from equilibrium give rise
to a set of new concepts, including transient order-enhanced phases \cite{Fausti2011}, pre-thermalized transient states \cite{Moeckel2008,Kollath2007,Essler2014,Babadi2015}, prethermalization at a quantum critical point \cite{Gagel2014}, and ordered current-carrying steady-states \cite{Prosen2012,Prosen2014,Ribeiro2014}.

A particularly interesting phenomenon arising in autonomous quantum systems, but yet poorly understood for open ones \cite{Lang2016,Sieberer2015,Marino2016}, is the occurrence of dynamical phase transitions. 
A dynamical transition is a singular point in the space of the couplings that parametrize the quenching protocol where an infinitesimal difference in parameters yields a qualitative difference for the asymptotic long-time state. 
First reported for quenches of integrable Bardeen-Cooper-Schrieffer model \citep{Barankov2004,Yuzbashyan2006,Yuzbashyan2006a,Barankov2006}, dynamical transitions were observed on a number of autonomous systems and studied using different methods \citep{Eckstein2009,Schiro2010,Sciolla2010,Schiro2011,Tsuji2013,Sciolla2013}. 
For open systems, the fundamental question of whether a dynamical transition can arise in the presence of an environment, remains unanswered. 
The coupling to a thermostat induces relaxation in the dynamics and subsequent decay to the equilibrium state. 
This process is expected to destroy the qualitative difference between the phases separated by the transition thus broadening it to a crossover. 

\begin{figure}[t!]
\center{\includegraphics[width=1\linewidth]{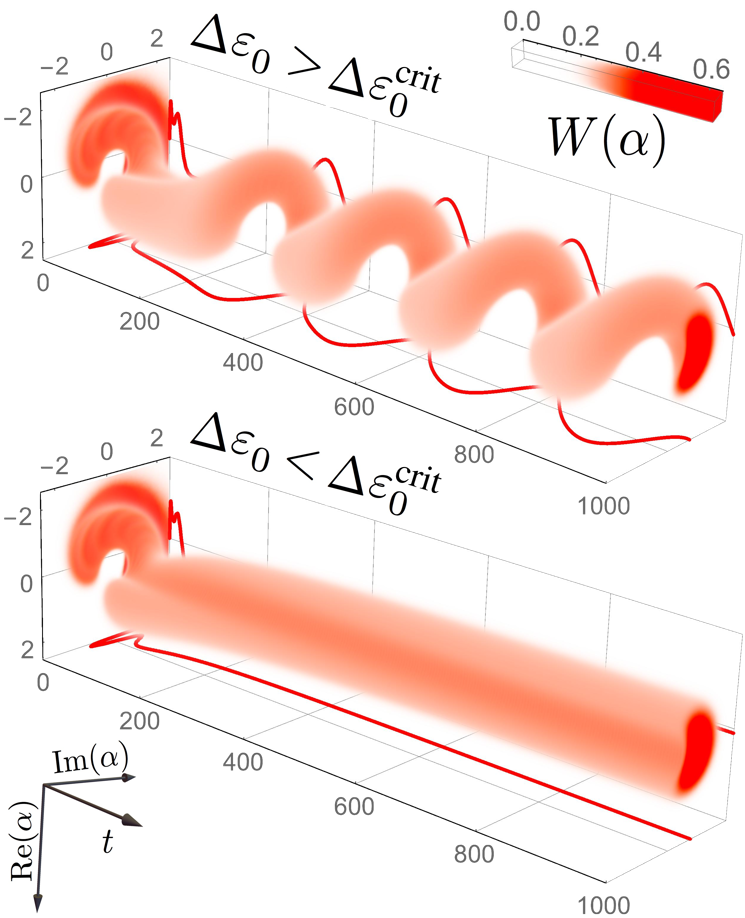}} 

\centering{}\caption{   
Post-quench time dependence of the Husimi function of the generalized Bose-Anderson model: $W= \left| \left\langle \alpha \right.\left| \Psi(t)\right\rangle 
 \right|^2/\av{\alpha|\alpha}$; $\ket{\alpha}=e^{\alpha a^\dag - \alpha^* a} \ket{0}$. The system was quenched between two points within the local Bose-Einstein condensed phase (D series in the Fig. \ref{Pdiagram}).
 Dependent on the value of the quench, the system evolves either to the new equilibrium state (lower panel) or 
 to the stable excited state excibiting a persistent phase rotation of the order parameter (upper panel). 
}
\label{Husimi} 
\end{figure}

In this paper we show that impurity models may be excited to a set of states that are completely decoupled from the environment, therefore avoiding relaxation and ensuring the existence of a dynamical transition. 
Fig. \ref{Husimi}  illustrates  the long time post-quench evolution of the order parameter. 
It shows the Husimi function of a localized interacting  bosonic mode coupled to a reservoir.  
The lower panel depicts the expected relaxation to the equilibrium state. 
One can observe that in the relaxational regime the phase of the order parameter stops rotating for long times.
The upper panel shows how the broken-symmetry phase develops when the magnitude of the quench is large. 
It is worth observing that in the latter case the phase of the order parameter remains rotating for long times. In our work we show that the system indeed exhibits frictionless rotation of the order parameter and that a decoupling from the thermostat is realized for a particular direction of the rotation (counter-clockwise).
We will demonstrate that the post-quench evolution yields one of these two situations: either the equilibrium state or the state with persistent phase oscillations, and that these two regimes are separated by a dynamical transition.

We address these questions by considering a particular example, the Bose-Anderson single-impurity model
(B-SIAM) \citep{Lee2007,Lee2010,Warnes2012}, describing a correlated impurity
coupled to an infinite number of non-interacting bosonic lattice modes.
Albeit their apparent simplicity, the physics of impurity models is extremely
rich. Perhaps the most paradigmatic example of an impurity system
is the celebrated Kondo model \citep{Kondo} that has permitted to
explain the resistance minimum of certain metallic compounds.
Some impurity systems undergo so-called impurity phase transitions \citep{Si2001,Vojta2006,Bayat2014}
where local response and correlation functions become singular and
can be characterized by a set of critical exponents. At these zero-temperature
transitions only a non-extensive term in the free energy becomes singular.
Examples of widely studied quantum critical models  are the pseudo-gap Kondo \citep{Ingersent2002},
the spin-boson \citep{Hur2010} of the two impurity Kondo model \citep{Ingersent2002}. 

Quenched dynamics in impurity models has been considered in the context
of electronic transport on quantum dots \citep{Meisner2009,Werner2010,Antipov2016, Shchadilova:2014aa, Ghosh:2015aa} and
impurities in the quantum gases \citep{Shashi2014, Knap2015,Schmidt2016}. For systems with impurity phase transitions
quenches have been studied for the pseudo-gap Kondo model \citep{Schiro2012,Schiro2010b}
and for the spin-boson problem \citep{Orth2013,Henriet2014}. 
However, whether dynamical transitions are to be seen in impurity models is an open question.
If present, they have to be of a different nature of their extended
counterparts. In the same way the existence of transient long-lived
phases is a possibility not yet explored for impurity systems.

The B-SIAM,  addressed in this work, is of particular interest as it possesses a rich phase diagram with a set of impurity phase transition lines. 
Recent progress in the manipulation of ultra-cold atomic gases rendered the controlled realization of this model at experimental reach using optical lattices with single-site resolution \citep{Ciz2010}. This technique allows to manipulate localized potentials in real time. 
Creating local correlated defects allows for the possibility of studying bosonic impurity-like systems,  analogous to magnetic impurities or quantum dots in solid state devices,
with the crucial advantage of being able to probe the dynamics in real time.

Equilibrium studies of the B-SIAM, by numerical renormalization group \citep{Lee2007,Lee2010}
and exact diagonalization \citep{Warnes2012}, revealed a zero-temperature
phase diagram containing: high-symmetry phases, where number of excitations
in the local node is an integer; and a broken-symmetry phase. As these
phases are local counterparts of the Mott and superfluid phases, observed
for example in the Bose-Hubbard model on a lattice \citep{Freericks-BH,Werner-BH}, they are dubbed
local Mott Insulator (lMI) and local Bose-Einstein condensate (lBEC) in the
following. The local spectral function was shown to behave in a power
law fashion near zero frequency, with a negative exponent in the lBEC
and a positive one in the lMI \citep{Lee2010} that depend
on the density of states of the bosonic environment.
To our knowledge there were no previous attempts to study this model
away from equilibrium. 

A number of works consider interaction quenches
in extended systems featuring Mott and superfluid phases such as the
Bose-Hubbard lattice model \citep{Kollath2007,Roux2009,Roux2010,Zhang2011,Snoek2011,Trefzger2011,Sorg2014,Kain2014,Strand2015}.
Here, as the energy injected into the system is extensive the relaxation
to the new ground state is forbidden by energy conservation. Instead,
the system was found to approach either a thermalized
or a non-thermalized state depending on the quench magnitude \citep{Kollath2007}, with a dynamical transition
separating the two regimes.
Here, again, the presence of a thermostat is expected to dramatically change this picture since it induces relaxation of both thermal and non-thermal states eventually transforming the dynamical transition to a crossover. 
Numerically resolving between a dynamical transition and crossover regime in quantum many-body systems might be a challenging task. 
This problem is particularly difficult as it requires numerically exact methods to achieve long-times. 
For example, real-time Quantum Monte Carlo solvers suffer from the sign (phase) problem
\cite{Antipov2016}, whereas DMRG \cite{Wolf2014} and exact-diagonalization \cite{Balzer2014} are limited by the dimension of the basis of wavefunctions used. 
Moreover, approximate schemes approaches such as  NCA  impurity solver \cite{Strand2015} also suffer from memory issues raising at the long-time scale.

In this paper we address the existence of dynamical transitions by
studying the dynamics after a quench of the Bose-Anderson
single-impurity model generalized to $N$ components.
The number of components $N$ controls lattice quantum fluctuations and ensues
an exact solution in the $N\to\infty$ limit \citep{Bickers1987} therefore 
allowing for the study the full non-equilibrium dynamics of the system with a small numerical effort.
We observe a variety of after-quench dynamical patterns, including relaxation to the equilibrium state, intermediate
states qualitatively different from the initial and final ones, the formation of so-called transient phases, and stable recurrent states. 
We provide numerical evidence and physical argumentation that these recurrent states are indeed completely decoupled from the lattice thermostat. 
Hence a dynamical transition is found, that separates the evolution from asymptoticly reaching the equilibrium or leading to stable recurrent states. 
Finally we argue that most of these effects hold for finite values of $N$ rendering our findings of immediate experimental relevance.

The structure of the paper is as follows. We present the model and methods in Sec. ~\ref{sec:2}. 
In Sec. ~\ref{sec:3} we expose our results, including the equilibrium phase diagram and the qualitative different quench types. 
We present an integrated discussion of the results and develop some analytic arguments in Sec. ~\ref{sec:4}.
We conclude in Sec. ~\ref{sec:5}. 
Appendix Sec. ~\ref{sec:LN} is devoted to the formal prove of the exactness of our method for large $N$.

\section{Model and method}
\label{sec:2}
\begin{figure}[t]
\begin{centering}\center{\includegraphics[width=0.8\linewidth]{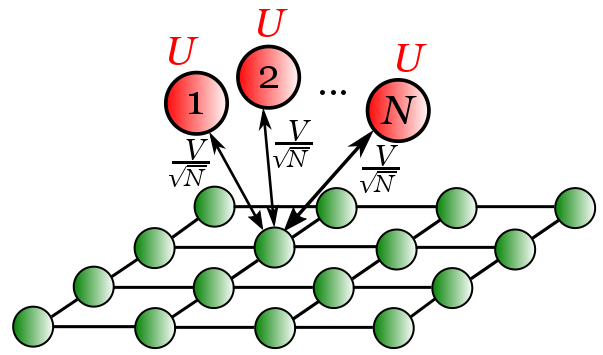}}
\par\end{centering}
\protect\caption{Sketch of the generalized Bose-Anderson model. $N$ impurities are coupled to the lattice at the same point with the same coupling constant $\frac{V}{\sqrt{N}}$. The on-site interaction $\frac{U}{2}n(n-1)$ occurs only at the impurities.}
\label{sketch}
\end{figure}

We consider an impurity with $N$ components, connected to a bosonic lattice on a single site, as shown in Fig. \ref{sketch}. 
This model is described by the generalized multi-component Bose-Anderson Hamiltonian:
\begin{equation}
\label{generalizedH}
H=\sum_j H_\text{SI} [a_j^{\dagger} a_j] 
-  \sum_{j,k} \frac{V_k}{\sqrt{N}} (a_j^{\dagger} b_k + h.c.) + \sum_k \epsilon_k b_k^{\dagger} b_k,
\end{equation}
where 
\begin{equation}
\label{Himp}
H_\text{SI}[a^{\dagger} a]=\varepsilon_0 a^\dagger a + \frac{U}{2} a^\dagger  a^\dagger a a 
\end{equation} 
is the Hamiltonian of a single component. The operator $a_j^{\dagger}$, with $j=1,...,N$, creates a  $j$-component boson  at the impurity site, 
$\varepsilon_0$   is the depth of the impurity on-site potential and $U$ is the local interaction strength. 
The operator $b_k^\dagger$ creates a boson in mode $k$, $\epsilon_k $ is the lattice dispersion  and $V_k/\sqrt{N}$ is the hopping amplitude 
between the bath and the impurity. In this paper we define the energy scale by the condition $\hbar=1$.
We assume that $\epsilon_k$ has a single minimum near $\epsilon_{k=0}=0$. The mode with $k=0$ is excluded from the calculations to get rid of the effects, related to bulk BEC \cite{Warnes2012}. 
In the following we consider that the dispersion relation of the bosonic bath is taken to be that of a 3-dimensional cubic lattice, $\epsilon_{k}=2h\left(3-\cos k_{x}-\cos k_{y}-\cos k_{z}\right)$, where $h$  is the hopping matrix element. We set $h=1$, $U=1$. As the impurity is coupled to a single lattice site $V_{k}=V$  is independent of $k$. The considered quench parameters are either the on-site energy $\varepsilon_0(t) = \theta(-t) \varepsilon_0(0)  +\theta(t) \varepsilon_0 $ or the impurity-bath coupling $V(t) = \theta(-t) V(0)  +\theta(t) V $. 
In the limit $N=1$ this model becomes the single-impurity Bose-Anderson model~\cite{Lee2007,Lee2010,Warnes2012}.
The rescaling in the hopping amplitudes is introduced in order to obtain a well defined large $N$ limit.

In the following we give a derivation of the equations ruling the dynamics of the model in the large $N$ limit,
assuming an equilibrium zero temperature initial state. 
An alternative derivation, using non-equilibrium Green's functions, valid also at finite $T$ is given in Appendix \ref{sec:LN}. 
For the $T=0$ case, we consider the separable ansatz for the wavefunction
\begin{equation}
\label{wf}
\ket{\Psi}=\ket{\Psi_\text{bath}} \otimes \ket{\Psi_\text{imp}},
\end{equation}
derive the equations of motion for both $\ket{\Psi_\text{bath}}$ and $\ket{\Psi_\text{imp}}$
and show this treatment becomes exact for $N \to \infty$.
For finite $N$ this ansatz can be seen as  mean-field-like approximation.

With the factorized wavefunction (\ref{wf}) two effective Hamiltonians  
$H^\text{eff}_\text{imp} \equiv \bra{\Psi_\text{bath}} H \ket{\Psi_\text{bath}}$ and 
$H^\text{eff}_\text{bath} \equiv \bra{\Psi_\text{imp}} H \ket{\Psi_\text{imp}}$ 
can be defined for the impurity and the bath respectively. 
The effective dynamics is given by the two Schr\"odinger equations,
\begin{equation}
\begin{split}
i \partial_t \ket{\Psi_\text{imp}} = H^\text{eff}_\text{imp} \ket{\Psi_\text{imp}}, \nonumber \\
i \partial_t \ket{\Psi_\text{bath}} = H^\text{eff}_\text{bath} \ket{\Psi_\text{bath}}. \nonumber
\end{split}
\end{equation}
coupled through the dependence of the effective Hamiltonians on the instantaneous mean values of the bath or impurity observables.

The impurity effective Hamiltonian, $H^\text{eff}_\text{imp}=\sum_j H^\text{eff}_\text{SI}[a_j^{\dagger} a_j]$, factorizes into a 
sum over independent components
\begin{equation}
\begin{split}
H^\text{eff}_\text{SI}[a^{\dagger} a]=\varepsilon_0 a^\dagger a +  \frac{U}{2} a^\dagger  a^\dagger a a - \lambda a^\dagger - \lambda^{*} a,
\end{split}
\label{Heff}
\end{equation}
where we have introduced the parameter
\begin{equation} \label{lambda}
\lambda= \sum_k \frac{V_k}{\sqrt{N}} \beta_k,
\end{equation}
with $\beta_k=\av{b_k}$. 

The bath effective Hamiltonian is given by
\begin{equation}
\begin{split}
H^\text{eff}_\text{bath}=
\sum_k \epsilon_k b_k^\dagger b_k 
-\sum_{k,j} \frac{V_k}{\sqrt{N}} \left(b_k^\dagger \av{a_j}+\av{a_j^\dagger} b_k \right).
\end{split}
\end{equation}
Note that, since all the components interact equally with the bath and we assume a component-symmetric initial condition, all $\av{a_j}$ 
are the same and thus  $\sum_j \av{a_j}=N\av{a}$. Similarly, other componentwise averages are independent of the component index: $\av{n_j}\equiv \av{a^\dag_j a_j} =\av{n}$, $\av{n_j^2}-\av{n_j}^2=\Delta^2 n$, etc. Thus further in the text we will appeal to the averages per component $\av{a}, \av{n}$.

The equation for the mean value of $b_k$ then reads
\begin{eqnarray}\label{bmotion}
-i\frac{d \beta_k}{dt} = \av{\left[H^\text{eff}_\text{bath}, b_k\right]} = -\epsilon_k \beta_k + V_k \sqrt{N} \av{a}.
\end{eqnarray}
which in its integral form yields 
\begin{eqnarray}
\beta_k (t) = \beta_k (0)  e^{-i \epsilon_k t} +  i V_k \sqrt{N} \int_{0}^{t} \av{a (t')} e^{-i \epsilon_k (t-t')} dt'.
\end{eqnarray}
With this expression the value of $\lambda$ in Eq.~(\ref{lambda}) becomes
\begin{equation}
\lambda(t) = \sum_k V_k^2 \frac{e^{-i \epsilon_k t}}{\epsilon_k} \av{a (0)} 
+ i  \sum_k V_k^2 \int_{0}^{t}  \av{a (t')} e^{-i \epsilon_k (t-t')} dt'.
\label{lambda_t}
\end{equation}

Eq.~(\ref{Heff}) together with (\ref{lambda_t}) form a closed set of equations describing the 
effective dynamics of the multicomponent Bose-Anderson Hamiltonian after a quench.
The initial condition is assumed to be the equilibrium solution before the quench is performed.
$\lambda(0)$ is computed solving the self-consistent condition obtained from 
Eqs.~(\ref{Heff},\ref{lambda_t}) at $t=0$.

Note that the considered ansatz wave function amounts, for this model, to consider the bath in the coherent state: 
\begin{equation}
\label{equilibrium}
\ket{\Psi_\text{bath}} = \Pi_{k} e^{\beta_k b^{\dagger}_k - \beta_k^* b_k} \ket{0}.
\end{equation}

Let us now argue on physical grounds why this ansatz becomes exact in the limit of infinite $N$.  
According to Eqs.~(\ref{bmotion}, \ref{equilibrium}) the displacement of lattice modes scales as 
$\av{b_k}\propto\sqrt{N}$.
For large $N$, this means that the lattice oscillators are in the classical regime and thus become c-numbers 
in the limit of infinite $N$. 
Each lattice mode is thus described by a single value $\av{b_k}$, which obeys the classical equation of motion (\ref{bmotion}). 
The evolution of the impurity can be found assuming a factorized form for the  wavefunction, 
$ \ket{\Psi_\text{imp}}=\ket{\Psi_\text{SI}}_1 \otimes ... \otimes \ket{\Psi_\text{SI}}_N$ 
that is exact once the bath modes are c-numbers. 
The wavefunction $\ket{\Psi_\text{SI}}$ obeys the Schr\"odinger equation with the single component Hamiltonian (\ref{Heff}).
Eq.~(\ref{Heff}) describes a single bosonic mode it is thus amenable to be treated numerically. 
The formal proof that the separable ansatz is exact in the limit $N\to\infty$ is given Appendix \ref{sec:LN}.

\section{Results}
\label{sec:3}

In this section first we present the equilibrium phase diagram of the system and describe the distinct phases.  In the following subsection we discuss the numerical results for the evolution of the system after quenches of the system parameter. The initial state is always taken to be the ground state of the starting phase. In all further calculations we consider time $t$  in units of inverse hopping energy $h^{-1}$.

\begin{figure}[t]
\center{\includegraphics[width=1\linewidth]{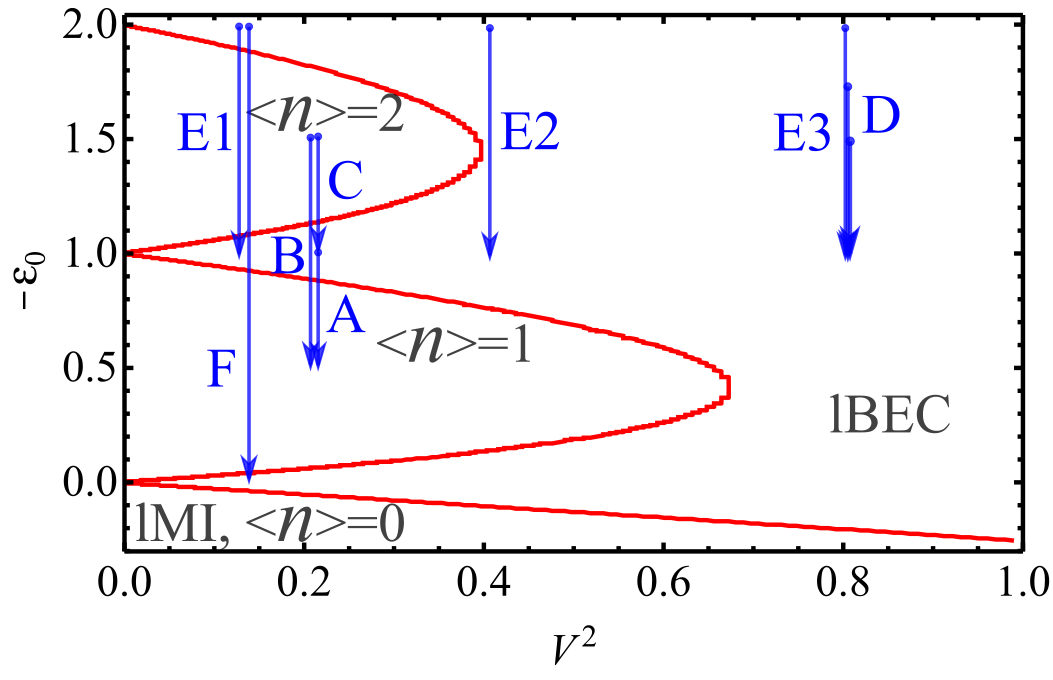}} 

\centering{}\caption{Equilibrium phase diagram of generalized Bose-Anderson model at $U=1, h=1$. Different
lobes correspond to a local Mott insulator phase with a different
number of bosons per mode. The local Bose-Einstein condensate is characterized
by a non-vanishing values of the order parameter. The blue arrows
depict the set of initial and final conditions characterizing the quench protocols considered in the paper.}
\label{Pdiagram} 
\end{figure}

\subsection{Equilibrium }

The equilibrium phase diagram of the generalized Bose-Anderson model
is shown in the Fig.~\ref{Pdiagram}. The results were obtained solving
the self-consistent condition of Eqs.~\eqref{Heff} and \eqref{lambda_t} at equilibrium (that is assuming all time derivatives in Eq.~\eqref{bmotion} are equal to zero). 
The phase diagram encompasses two phases: a set of
lMI lobes with a vanishing order parameter $\av{a}$, and a lBEC phase where $\av{a}\neq0$.
In the lBEC a finite value of $\av{a}$ implies non-vanishing fluctuations of the number of particles at impurity site, $\Delta^2 n > 0$.
Moreover, the number of particles $\av{n}$  is not restricted to be an integer.  
The number of particles on the lattice is given by the expression $\sum_k \av {b_k^\dag b_k}=N\sum_k V_k^2 \frac{|\av{a}|^2}{\epsilon_k^2}$. 
For a $d$-dimensional lattice with $d<4$ at $V_k=V$, the sum diverges in the thermodynamic limit whenever $\av{a} \neq 0$, thus indicating the formation of a BEC. Note that the bulk of the lattice lacks the BEC mode $k=0$ since for convenience it is not considered here, therefore $\sum_k \av {b_k^\dag b_k} \to \infty$ signals the formation of the BEC cloud in the vicinity of the impurity.
The lMI phase is characterized by an integer number of bosons at the impurity
site, i.e. $\av{n} \in\mathbb{N}_{0}^{+}$ and $\Delta^2 n = 0$. 
Since $\av{a}=0$, lattice modes are not populated in this phase.

The phase diagram is qualitatively
similar to the single-impurity Bose-Anderson model as obtained by
numerical renormalization group approach \cite{Lee2007,Lee2010} and by the exact diagonalization method \cite{Warnes2012}. 
The nomenclature of the phases
is reminiscent of the Bose-Hubbard model \cite{Freericks-BH,Werner-BH}
equilibrium phase diagrams.

In the following subsections we consider the dynamics of the system after the instantaneous change of the chemical potential on the impurity site. Fig. \ref{Pdiagram} depicts the set of different quenches under consideration. 
Quench protocols are characterized by the initial and final conditions. For each quench we study the time evolution of the local order parameter $\av{a}$ and of the number of particles per mode at the impurity site.

\subsection{Quenches to the lMI phase}

\paragraph*{\textbf{lBEC-lMI quench}.--}

\begin{figure}[t]
\center{\includegraphics[width=1\linewidth]{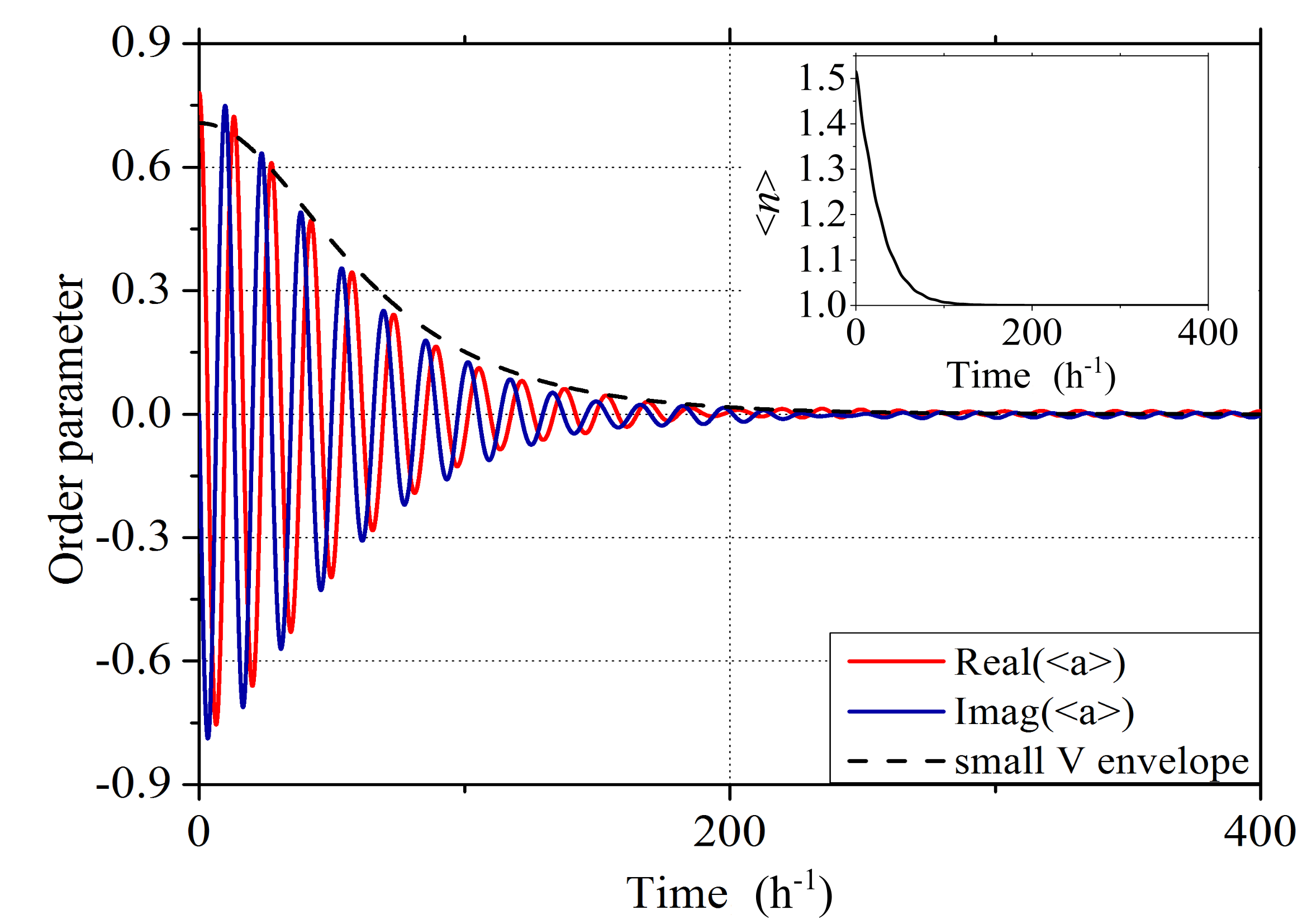}} 
\centering{}\caption{Time evolution of the local order parameter $\av{a(t)}$, real and imaginary part, after the quench 
of the local energy on the impurity site from the lBEC to the lMI phase with $\av{n}=1$.
Parameters of the quench are: $\varepsilon_{0}(t=0)=-1$ and $\varepsilon_{0}(t> 0)=-0.5$ at fixed coupling $V=0.447$  
(this quench protocol is schematically depicted with the arrow A in Fig.~\ref{Pdiagram}).
Dashed line shows the envelope calculated analytically in the weak coupling limit using Eq.~\eqref{envelope}.
Time evolution of the average number of particles per mode at the impurity site $\av{n(t)}$ for the same quench parameters is shown in the inset.}
\label{SFMIquench} 
\end{figure}

First of all let us consider the evolution of the system in the lBEC phase
after the instantaneous change of the local energy on the impurity site.
We choose the final value of the local chemical potential such as the final state 
corresponds to the lMI phase at the equilibrium phase diagram. In Fig.~\ref{SFMIquench} the evolution 
of the local order parameter is shown after the quench which is schematically depicted 
with arrow A on the equilibrium phase diagram in Fig.~\ref{Pdiagram}. 
Evolution of the order parameter shows fast oscillations which decay in the long time limit. 
The evolution of the average number of particles at the impurity $\av{n(t)}$ 
is shown in the inset in the Fig. \ref{SFMIquench}. The average number of particles decays 
to the expectation value defined by the final Hamiltonian.

The fast oscillatory behavior of the order parameter is related to the internal impurity dynamics.
The impurity Hamiltonian (\ref{Himp}) after the quench 
of the local energy on the impurity site to the value $\varepsilon_0=-0.5$
has an energy gap between the ground state and the first excited state equal to $E_{1,0}= 0.5$.
The period of the fast oscillations of the order parameter roughly corresponds to $2 \pi/E_{10}\approx 12.6$. 
The slight variation of the period of oscillations can be attributed to the effect of the lattice degrees of freedom on the impurity dynamics. The fast oscillations of the order parameter appear for all quenches when the coupling parameter is small, $V^2/(U h)\lesssim 0.4$.

The time scale associated with the slow decay is calculated analytically in the limit of the weak coupling and is shown in Fig.~\ref{SFMIquench}. It shows good qualitative agreement and its derivation is shown in Section~\ref{tr_ph}.

\paragraph*{\textbf{lMI-lMI quench}.--}
The evolution of the local order parameter after the quench between two lMI 
phases (quench protocol B) with different number of particles at the impurity site is considered.
In the initial lMI state the equilibrium value of the order parameter is zero, $\av{a(t=0)}=0$.
Such a state is a fixed point of the equations Eqs.~(\ref{Heff},\edit{\ref{bmotion}},\ref{lambda_t}). To allow the nontrivial dynamics of the order parameter $\av{a(t)}$ starting from the lMI phase 
a small deviation of the order parameter from its vanishing equilibrium value is assumed, $\av{a(t=0)}=\delta a$.
We call the value $\delta a$ as a seed noise.
The presence of symmetry breaking fluctuations slightly shifts the system away from its unstable fixed point 
and allows for the transition between two symmetric phases. 

\begin{figure}[t]
\center{\includegraphics[width=1\linewidth]{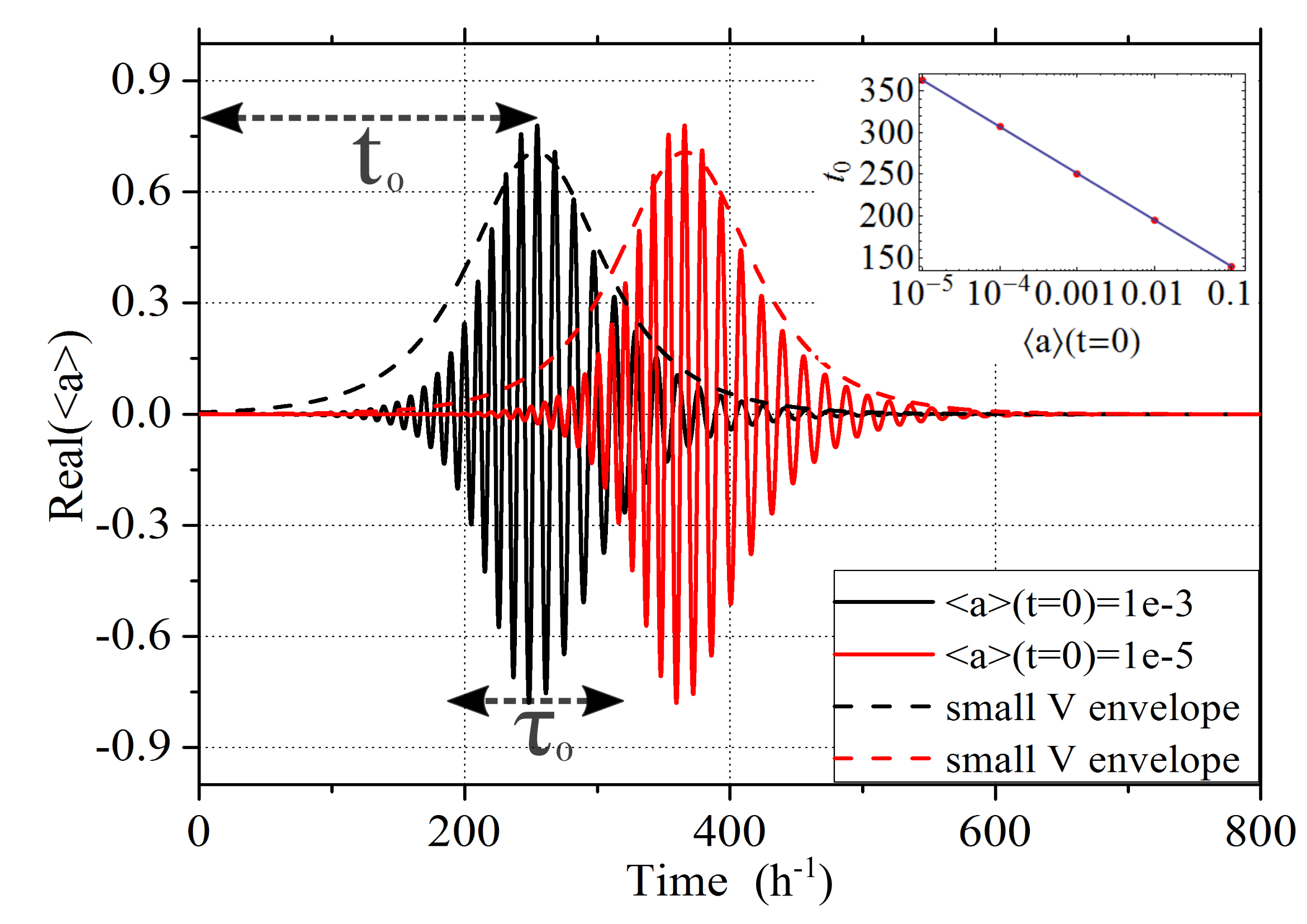}} 
\caption{Time evolution of the real part of the local order parameter $\av{a(t)}$ after the quench of the local energy on the impurity site
from lMI phase with $\av{n}=2$ to lMI phase with $\av{n}=1$ is shown for different values of the initial seed noise $\av{a(0)}$.
Parameters of the quench: $\varepsilon_{0}(t=0)=-1.5$ and $ \varepsilon_{0}(t>0)=-0.5$ with $V=0.447$
(this quench protocol is schematically depicted with the arrow B in Fig.~\ref{Pdiagram}).
The lifetime of the intermediate phase is $\tau_0$.
Dashed line shows the envelope calculated analytically in the weak coupling limit using Eq.~\eqref{envelope}.
The inset shows the dependence of the response time $t_0$ on the initial seed noise amplitude $\av{a(0)}$. }
\label{MIMIquench} 
\end{figure}

Fig. \ref{MIMIquench} shows the post-quench evolution of the local order parameter. 
After the quench the order parameter grows from its initial
value defined by the seed noise $\delta a$.
Similar to the lBEC-lMI transition, the period of fast oscillations of the order parameter is determined by energy differences of the isolated impurity Hamiltonian.
At certain time $t_{0}$ the amplitude of oscillations reaches its maximum.
This time decreases with the amplitude of
the seed noise as $ t_{0}\propto - \mathrm{log}(\delta a)$, the numerical result is shown in the inset in the
Fig. \ref{MIMIquench}. After the maximal amplitude is reached, the oscillations decay. Finally, the new ground state is
approached and the average number of particles saturates at the value defined by the phase of the final Hamiltonian. 
In this way the system undergoes a transition between lMI lobes
with different number of particles through a transient lBEC phase.

In the discussion section \ref{tr_ph} we show that the observed relaxation to the lMI state can be described analytically
assuming a weak coupling between the impurity and the lattice. Such an approach provide a good estimation for the envelopes of the time dependence of the order parameter. Their ``slow'' timescale $\tau_0$ is determined by the coupling constant and the density of the lattice states. 

\subsection{Quenches to lBEC phase} \label{tolBEC}
\begin{figure}[t]
\center{\includegraphics[width=1\linewidth]{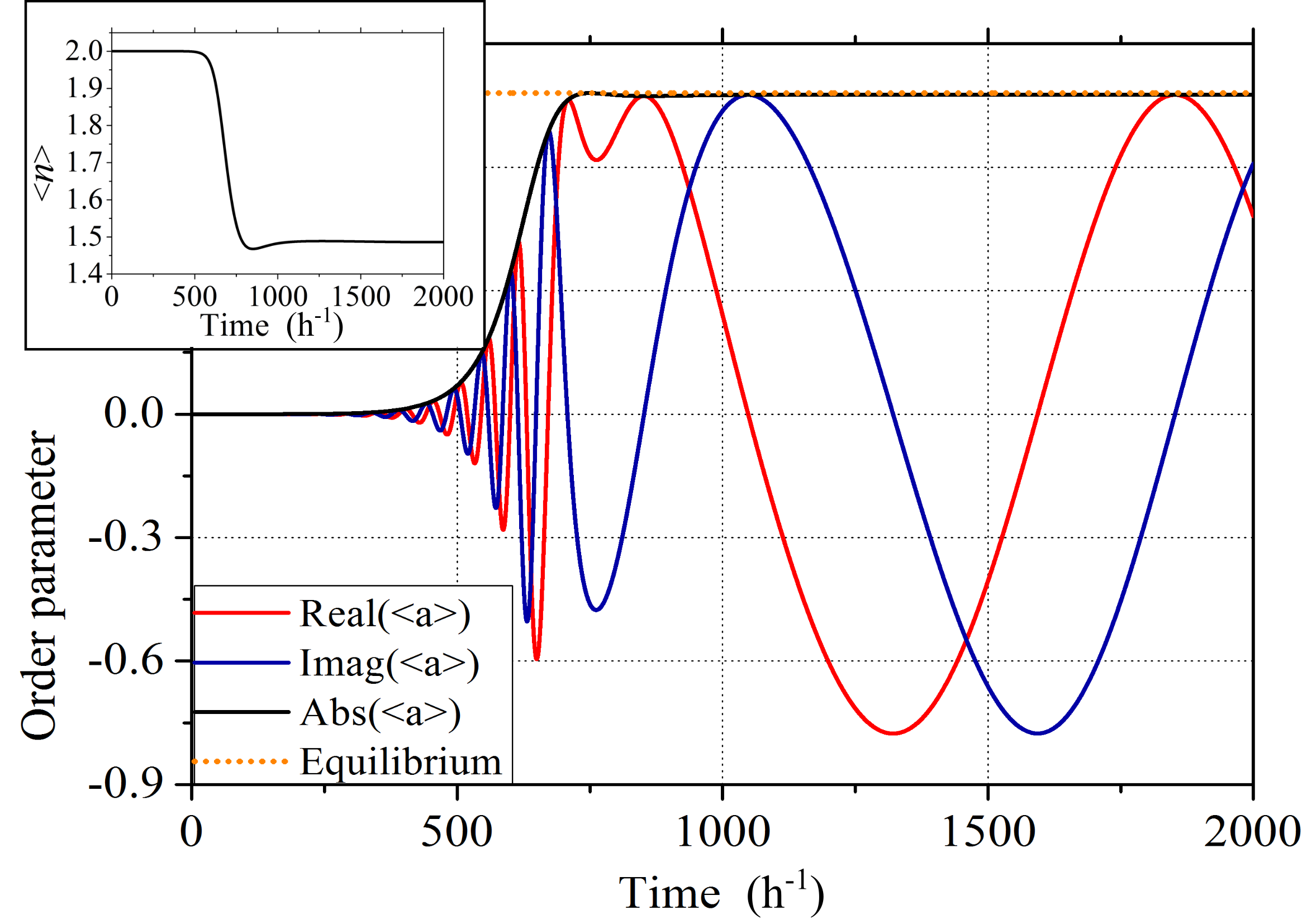}} 
\centering{}\caption{
Evolution of the local order parameter $\av{a(t)}$ -- its real and imaginary part, as well as its absolute value -- after the quench of the local energy on the impurity site from the lMI phase with $\av{n}=2$ to the lBEC phase is shown.
Parameters of the quench: $\varepsilon_{0}(t=0)=-1.5$ and $ \varepsilon_{0}(t>0)=-1$ with $V=0.447$
(this quench protocol is schematically depicted with the arrow C in Fig.~\ref{Pdiagram}).
Expectation value of the order parameter in the ground state of the final Hamiltonian is shown with the dashed line.
Inset shows the evolution of the average number of particles at the impurity site $\av{n(t)}$.}
\label{MISFquench} 
\end{figure}

\paragraph*{\textbf{lMI-lBEC quench}.--}
We now consider a quench from the lMI phase to the lBEC phase. 
Time evolution of the order parameter after the quench of the local energy is presented in the Fig. \ref{MISFquench}.  
Parameters of this quench are schematically depicted 
with arrow C on the equilibrium phase diagram in Fig.~\ref{Pdiagram}.
As in the case for lMI-lMI quench the initial state is an unstable stationary
point of the equations of motion of the order parameter. 
A small seed noise has been introduced into the initial conditions
to allow a nontrivial dynamics of the order parameter with possible transition between the two phases. 
At short times there is a rapid increase of the amplitude of the order parameter 
oscillations. The amplitude saturates after a time period dependent on the initial seed noise amplitude, similar to lMI to lMI quench considered above. 
For longer times the phase of the order parameter
rotates persistently  and the order parameter never reaches its steady-state. 
The number of particles at the impurity site decreases from the quantized initial  value and saturates
to the new equilibrium state, see inset in the Fig. \ref{MISFquench}. 
Calculations for various initial values of the seed noise show that  the persistent oscillations of the order parameter are the robust feature of the system, although the period of the oscillations varies with the particular choice of the seed noise. 

The phase of the order parameter shows non-trivial features in its dynamics.  
The order parameter vector rotates on the complex plane clockwise up to certain moment when the direction of the rotation changes to the counterclockwise one. 
This means that the rotation frequency interpolates between a positive value for short times and negative constant value for asymptoticly
large times. For the dynamics shown in the Fig. \ref{MISFquench} the transition between the two regimes takes place around $t\simeq 750$.

\paragraph*{\textbf{lBEC-lBEC quench}.--}
We consider the quenches of the local energy when the parameters 
of the initial and final Hamiltonian are chosen to correspond to the system in the lBEC phase.
On the equilibrium phase diagram we depict these quenches as D, E,and F, see  Fig. \ref{Pdiagram}.
There are two distinct cases: (i) quenches within the lBEC phase without crossing
any lMI lobe (e.g. quenches D in Fig.~\ref{Pdiagram}) and (ii) quenches that crosses at least one lMI lobe, e.g. quenches E and F in Fig. \ref{Pdiagram} cross one lMI lobe or two lMI lobes correspondingly.

\begin{figure}[t]
\center{\includegraphics[width=1\linewidth]{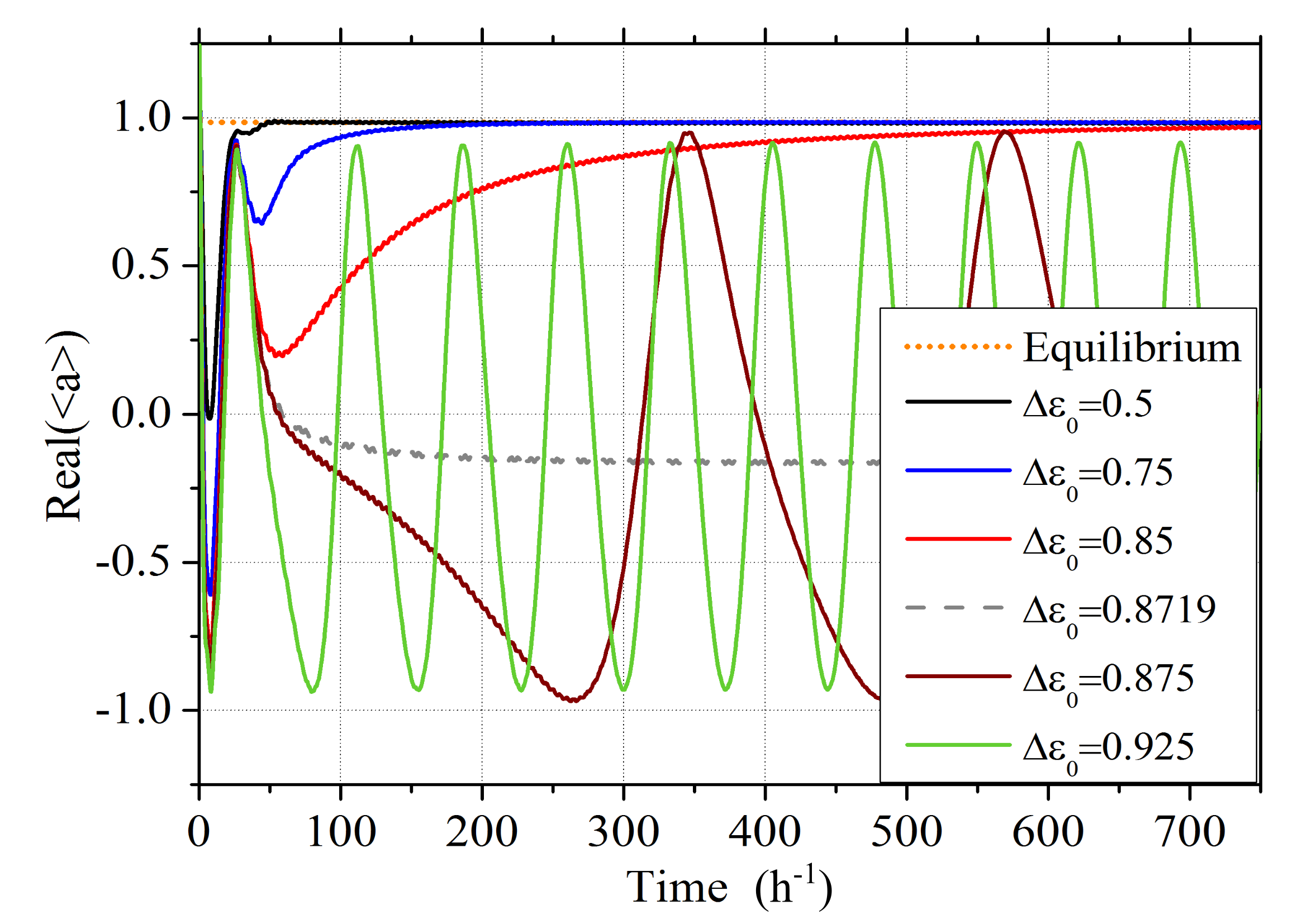}} 
\centering{}\caption{Time evolution of the real part of order parameter for quenches within the lBEC phase for different quench
amplitudes $\Delta\varepsilon_{0}=-(\varepsilon_{0}(t=0)-\varepsilon_{0}(t>0))$
for the same final value of the local energy $\varepsilon_{0}(t>0)=-1$ when interaction strength is fixed $V=0.89$ (these quenches are depicted with arrows D in Fig. \ref{Pdiagram}). 
There are two distinct behavior in the long time regime: (i) an equilibrated static steady-state
for small values of $\Delta\varepsilon_{0}$ and (ii) an oscillatory regime
with frequency and amplitude that depend on $\Delta\varepsilon_{0}$. 
Expectation value of the order parameter in the ground state of the final Hamiltonian is shown with the dashed line.}
\label{deltaeps} 
\end{figure}

First let us consider the quenches that do not cross any lMI lobe. 
We fix the final value of the local energy $\varepsilon_{0}(t>0)$ and vary the quench amplitudes
$\Delta\varepsilon_{0}=-(\varepsilon_{0}(t=0)-\varepsilon_{0}(t>0))$ by setting the initial condition.
As an example we consider a set of quenches depicted with D arrows in Fig. \ref{Pdiagram}. 
The post-quench time evolution of the real part of the order parameter is shown in the Fig. \ref{deltaeps}. 
For small values of the quench amplitude $\Delta\varepsilon_{0}$ the order parameter
approaches the new equilibrium value defined by the expectation value in the ground state of the final Hamiltonian. 
As $\Delta\varepsilon_{0}$ increases the long time evolution of the order parameter changes qualitatively 
from the relaxation to an oscillatory regime. 
A critical value of the quenching amplitude $\Delta\varepsilon_{0}^{crit}\simeq 0.8719$ separates 
these two very distinct regimes. At this singular point the solution is neither oscillating
nor reaching its static equilibrium value. As small deviations to this value lead to different asymptotic dynamical phases 
we will refer to this singular point as a \emph{dynamical phase transition}. 
Our numerical calculations show that the different asymptotic behavior is observed within the very small range of the quenching amplitudes close to the critical quench value, namely for $\Delta \varepsilon_0 = \Delta\varepsilon_{0}^{crit} \pm 10^{-4}$. 
This gives an evidence that we observe a true dynamical transition. 
Note that in the oscillatory regime both the amplitude and the frequency
of the oscillations depend on $\Delta\varepsilon_{0}$. As $\Delta\varepsilon_{0}$
tends to the critical value the amplitude approaches the
equilibrium static value and the frequency vanishes. For different final points of $\varepsilon_0 (t>0)$ our results (not shown) also demonstrate the presence of a dynamical transition.
However, the critical value of the quench amplitude $\Delta \varepsilon_0$ is different and the critical amplitude increases with $-\varepsilon_0(t>0)$ and $V$. 
Indeed, the dynamical transition can most easily be observed near the lMI-to-lBEC transition line.

\begin{figure}[t!]
\center{\includegraphics[width=1\linewidth]{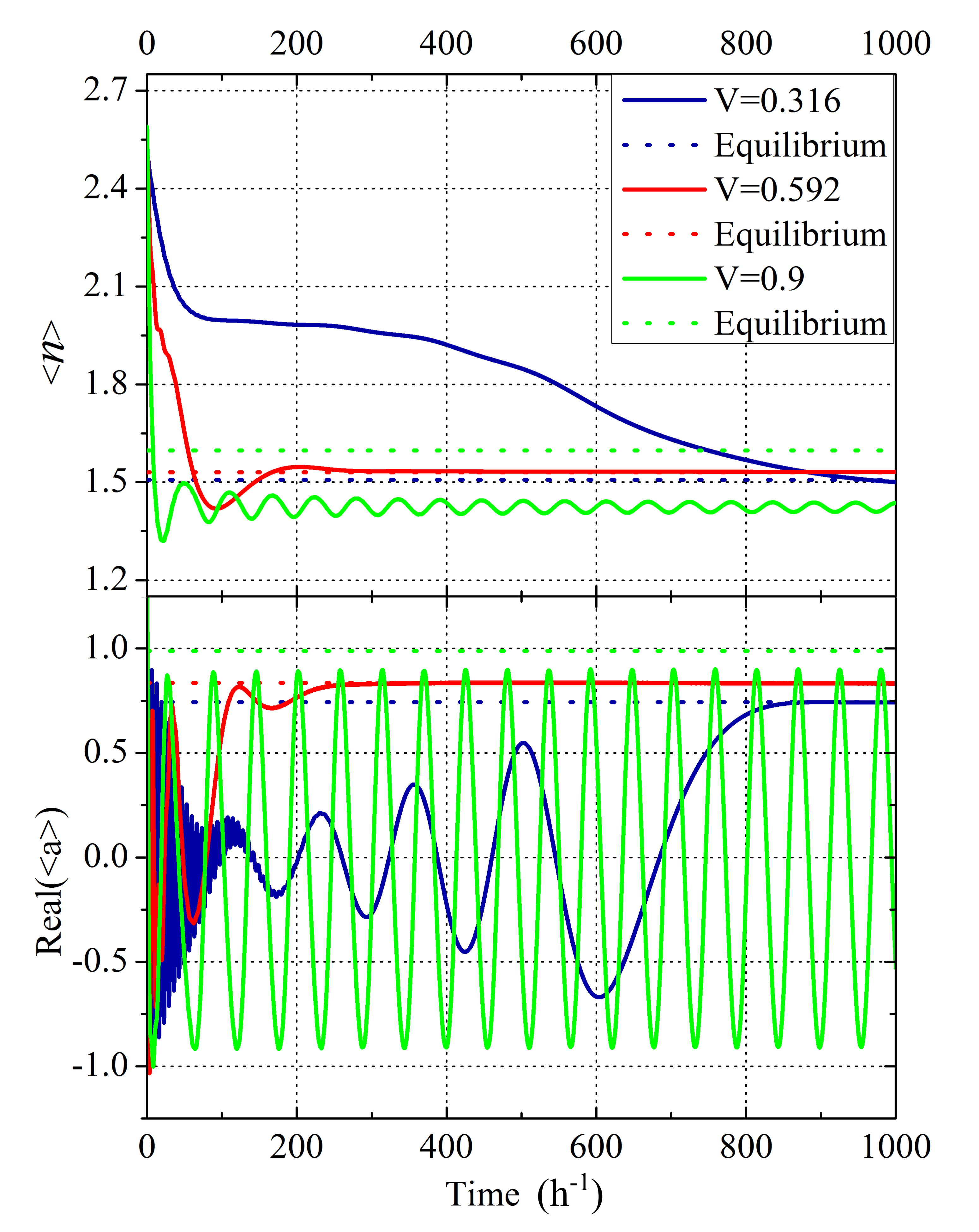}}
\center{\includegraphics[width=1\linewidth]{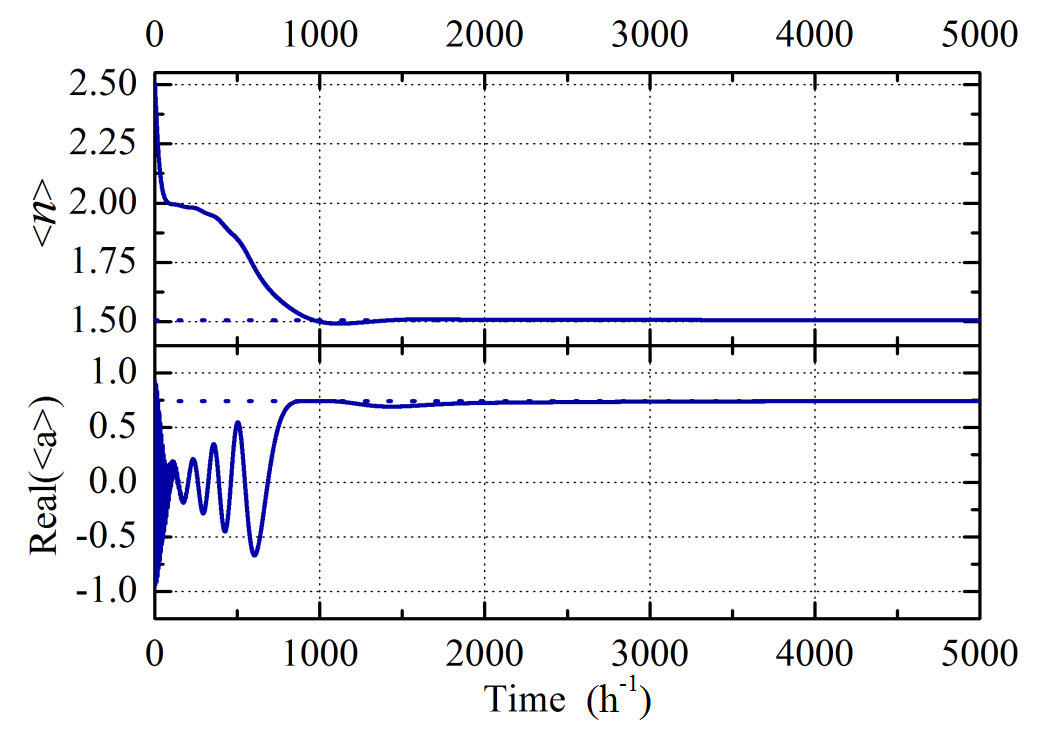}} 
\centering{}\caption{
Time evolution of the real part of the order parameter (upper panel) and number of particles on the impurity site $\av{n}(t)$ (middle panel)  
for fixed initial and final values of the local energy $\varepsilon_{0}(t=0)=-2$ and $\varepsilon_{0}(t>0)=-1$ correspondingly and 
three values of the coupling strength: $V=0.316$, $V=0.592$, and $V=0.9$.
These quenches are depicted with arrows E1-E3 in Fig. \ref{Pdiagram}.
Doted lines correspond to the expectation values $\av{a}$ and $\av{n}$ calculated in the ground state of the final Hamiltonian.
A plateau can be observed in the time evolution of the $\av{n}$ corresponding to $\av{n}=2$ in the case when lMI phases crossed during the quench. The long-time behavior of the order parameter and number of particles for the quench with $V=0.316$ is shown in the lower panel.
}
\vspace{-5pt}
\label{n_av_diff_v} 
\end{figure}

Finally we consider quenches within the lBEC phase that cross one or more lMI lobes, depicted with the arrows E1 and F in Fig. \ref{Pdiagram}. 
To highlight features of lMI lobe crossing observed in the evolution of the order parameter we compare case E1 with the quenches with the same initial and final values of the local energy and different value of the coupling strength $V$ such that no lobe is crossed 
(cases E2 and E3 in Fig. \ref{Pdiagram}).
Fig. \ref{n_av_diff_v} shows the time evolution of the real part of the order parameter and the average number of particles $\av{n(t)}$ 
for the three values of coupling strength $V$.
For the lobe crossing case, $V=0.316$, there is a plateau in the time evolution of the number of particles which corresponds to  $\av{n}\simeq 2$, this is the value of $\av{n}$ of the lMI lobe crossed.
In the long time limit the new equilibrium value is reached, defined by the final Hamiltonian.
When the interaction strength $V$ is large and no lMI lobe is crossed, there are two effects:
first the long-lived transient phase is lost, and second, for sufficiently large
interactions $V$, a dynamical phase transition to an oscillatory phase takes place. 
The latter behavior is similar to the one shown in Fig. \ref{deltaeps}.

\begin{figure}[t]
\center{\includegraphics[width=1\linewidth]{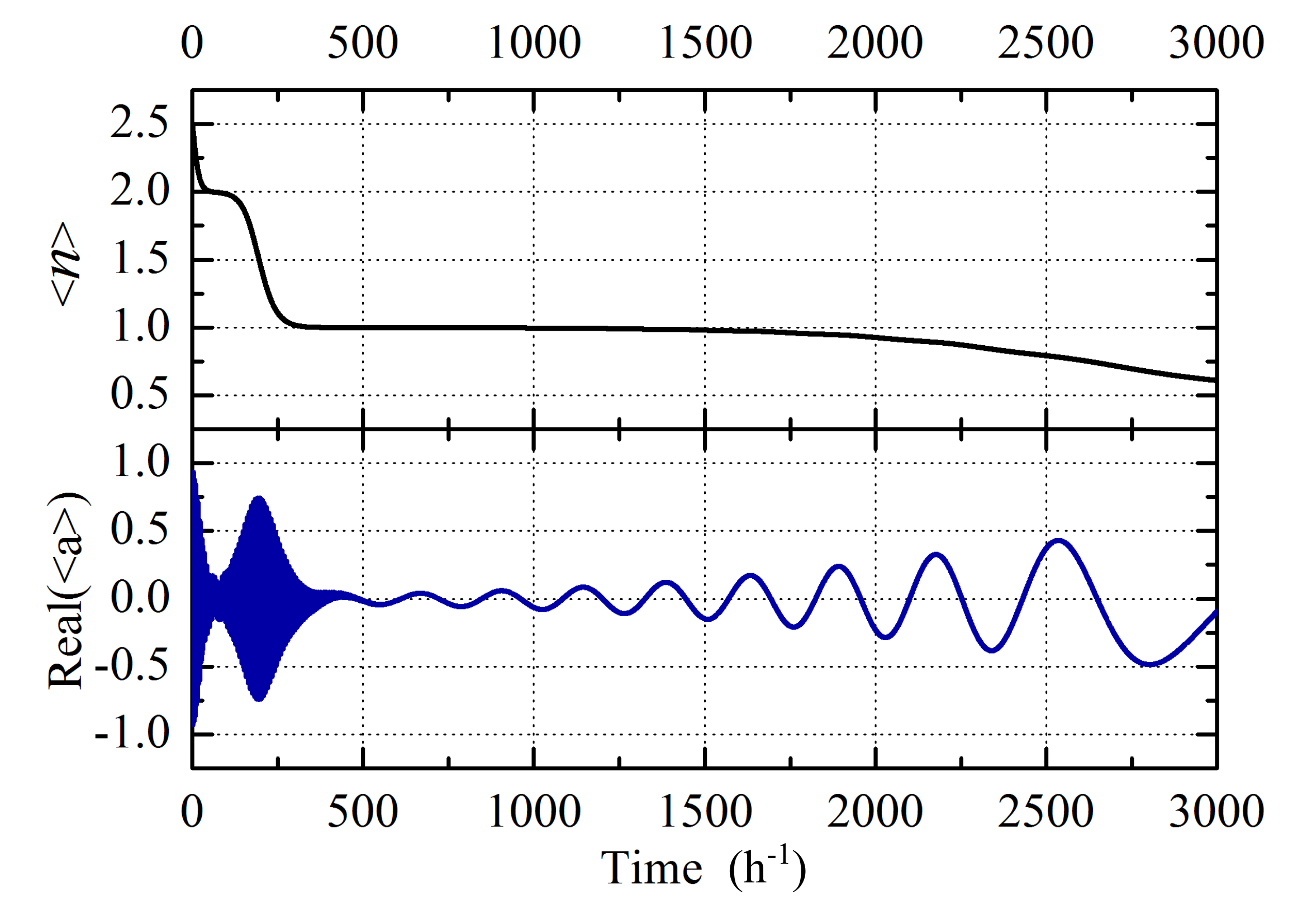}} 
\centering{}\caption{Time evolution of the real part of order parameter (down) and of number of particles (up) for the quench, depicted with arrow F, which cross two Mott lobes: $\varepsilon_0 (t=0)=-2, \varepsilon_0 (t>0)=0$ at $V=0.316$ .  
}
\label{cross2Mott} 
\end{figure}

The existence of a plateau with $\av{n}\simeq 2$ suggests that the long-lived transient phase acquires characteristic
features of the crossed lMI phase.
The phenomena of the locking of the particle number at an integer value whenever the corresponding lMI lobe is crossed is ubiquitous for this model. 
Fig. \ref{cross2Mott} shows evolution of the number of particles and of the order parameter after a quench crossing two lMI lobes. 
Two long-lived plateaus are observed in $\av{n}(t)$ around  $\av{n} \simeq 2$ and $\av{n} \simeq 1$ before the final relaxation to the local equilibrium value occurs. 
Within each plateau the amplitude of the oscillations of $\av{a}$ is small. In this way the system seems to mimic a vanishing order parameter during these long-lived transient phases. 

\subsection{Inverse quenches}

For all the quenches shown in Fig. \ref{Pdiagram} and described above the number of particles at the impurity site decreases with time with respect to its initial value. This process is always possible since there is an infinite number of bath modes to which particles can be emitted.  
We also considered quenches for which the equilibrium state of the final phase contains a larger number of particles at the impurity than the initial state does. 
It corresponds to the inversion of the direction of arrows  in Fig. \ref{Pdiagram}. 
Local equilibration is in this case never achieved as the bath effectively decouples. Indeed, 
the Hamiltonian \ref{generalizedH} conserves total number of particles in the system.
Bulk of the lattice contains no particles at the initial state of the system, therefore the bath is not able to provide extra particles for the impurity to achieve its local equilibrium state.    For completeness we provide a description of the time evolution in this case.
When starting from lMI the system seems to be locked in its initial phase - the order parameter amplitude undergoes an oscillatory motion with an amplitude that never exceeds the initially introduced seed noise. Thus contrarily to the previous case the initial state is stable and the system never leaves the initial lMI phase. 
When starting from lBEC phase the long time behavior is always oscillatory even for very small quench amplitudes. We have also verified numerically that quenches with zero amplitude don't change a state of the system, even in the presence of the seed noise.

\section{Discussion}
\label{sec:4}

\subsection{Transient phases} \label{tr_ph}

An interesting and striking feature of the post-quench dynamics is the appearance of long-lived transient regimes, when an equilibrium phase is crossed,
with the same characteristics of an underlying phase. 
We dubbed such regime a transient phase.

A transient lBEC phase can be observed during the transition regime between lMI lobes in the lMI to lMI quenches of Fig. \ref{MIMIquench}.  
During this regime the order parameter becomes finite and the number of particles fluctuates. 
The phenomena is even more striking in lBEC to lBEC quenches that cross one of more lMI lobes.
The long plateaus observed at integer values in the dynamics of $\av{n}(t)$ (see Fig. \ref{n_av_diff_v}) can be interpreted as the formation of  transient lMI phases.

A simple physical picture capturing the main features of lMI to lMI quenches can be obtained in the limit of small $V$.
This regime is usually considered in quantum optics, where atoms in excited states are coupled to the electromagnetic continuum \cite{Andreev}. 
Here we adapt the standard argument for the calculation of the super-radiance intensity to our system. 
For clarity, we consider only lMI to lMI transitions between phases with $n+1$ and $n$ particles. 
The weak coupling formally corresponds to a small value of $V$ and, consequently, of $\lambda$. 
We thus consider a perturbative series in this parameter. 

At zeroth order, the bath is in its vacuum state $\ket{0_\text{bath}}$, and
the wavefunction of an impurity is considered to be in a symmetric state with each mode in a superposition of  $n$ and $n+1$ particles:
\begin{equation}
\label{wflMI}
\ket{\Psi^{(0)}}=\ket{0_{bath}} \otimes_{j=1}^N \left(\phi_n \ket{n}_j+\phi_{n+1}\ket{n+1}_j\right).
\end{equation}
After a time $T$ the first order correction, corresponding to the irradiation of a particle to the bath, is given by the formula
\begin{equation}
\ket{\Psi^{(1)}}=\sum_{jk} b^\dag_k \tilde{a}_j \int_0^T e^{-i (\epsilon_k-\varepsilon_{n+1}+\varepsilon_n) t} dt \ket{\Psi^{(0)}}.
\end{equation}
Here $\tilde{a}= P a P$ is the annihilation operator projected into the considered subspace with $P = \ket{n+1}\bra{n+1} + \ket{n}\bra{n}$ and $\varepsilon_n$ is the energy of the impurity Hamiltonian with $n$ particles. 
The overall transition rate equals $W=\partial_T \av{\Psi^{(1)} \left| \Psi^{(1)}\right.}$.
Taking the $T\to \infty$ limit we can write $W = \lim_{T\to\infty}  T^{-1}  \av{\Psi^{(1)} \left| \Psi^{(1)}\right.}$.
Calculating the averages
\begin{equation}
\label{mjj}
\begin{array}{l}
\bra{\Psi^{(0)}} \tilde{a}^\dag_j \tilde{a}_j \ket{\Psi^{(0)}}=(n+1) |\phi_{n+1}|^2, \\
\bra{\Psi^{(0)}} \tilde{a}^\dag_j \tilde{a}_{j'} \ket{\Psi^{(0)}}=(n+1) |\phi_{n} \phi_{n+1}|^2, ~~j'\neq j.
\end{array}
\end{equation}
we obtain
\begin{equation}
W=2\pi (n+1) V^2 A_{\epsilon_{n+1}-\epsilon_{n}} \left( |\phi_{n+1}|^2 +(N-1) |\phi_n \phi_{n+1}|^2\right),
\label{eq:decay}
\end{equation}
where  $A_{\epsilon_{n+1}-\epsilon_{n}}$ is the normalized density of lattice states at the transition energy.
The first term in the parentheses describes the spontaneous emission yielding Fermi's golden rule. 
In the large $N$ limit the contribution of this term is small. 
The second term describes the super-radiance of synchronized impurity modes.

The quantity $W$ can be interpreted as a decay rate of the population of the state $\ket{n+1}$.
In the large $N$ limit we can thus write
$W=N \frac{d |\phi_{n+1}|^2}{d t}$, which yields the detailed balance equation
\begin{equation}\label{envelope0}
\left.\frac{d |\phi_{n+1}|^2}{d t}\right|_{N\to \infty}=- \frac{|\phi_{n+1}|^2 (1-|\phi_{n+1}|^2)}{\tau_0}.
\end{equation}
Here we introduced the timescale $\tau_0^{-1}={2\pi (n+1) V^2 A_{\epsilon_{n+1}-\epsilon_{n}}}$ and took into account the normalization condition $|\phi_{n}|^2+|\phi_{n+1}|^2=1$.
Solving the differential equation (\ref{envelope0}) we obtain the envelope of the order parameter in the weak-coupling limit:
\begin{equation}
\left|\av{a}\right|(t)=\sqrt{\frac{n+1}{2+2 \rm{ch} (t-t_0)/\tau_0}}.
\label{envelope}
\end{equation}
where $t_0$ is fixed by the initial condition. 
The instanton-like envelope functions obtained in this way are depicted in Fig. \ref{MIMIquench} and are in good agreement with the numerical results. The logarithmic dependence of $t_0$ on $\delta a$, shown in the inset of Fig. \ref{MIMIquench}, also arises from Eq. (\ref{envelope}).

A similar argument could be made for lBEC to lBEC quenchs at small $V$, shown in Fig. \ref{n_av_diff_v}.
Since three states, $\ket{n-1},\ket{n},\ket{n+1}$, are involved in the process a simple analytical solution could not be found.  
Nonetheless we can use the previous case to argue that this process can be seen as two half instanton-like solutions. 
After the quench, the impurities relax from the mixed state $\phi_3 \ket{3}+\phi_2 \ket{2}$ with approximately equal $\phi_{2,3}$.
This stage of the evolution at $t\approx 100$ yields almost a pure lMI $n=2$ state; the value of
order parameter is decreased considerably. 
However, this transient lMI phase is not stable and decays to lBEC. 
The period of the long-time oscillations of the order parameter seems to vary with time suggesting an intricate interplay of degrees of freedom in the course of the evolution.

In this paper we limit our study to the $N \to \infty$ case. 
Nonetheless, possible realizations of the generalized multi-component Bose-Anderson Hamiltonian forcefully feature a finite number of modes. Thus, we now discuss experimentally relevant finite-$N$ effects. 
Qualitatively two relaxation mechanisms are possible, as can be seen from Eq. (\ref{eq:decay}): the collective super-radiance, discussed above, and a set of independent spontaneous emission processes. 
The latter are characterized by the single particle lifetime $t_{sp}=\tau_{0}N$, that is $N$ times larger than the lifetime due to super-radiant processes. 
Consequently the relative impact of spontaneous emission processes can be roughly estimated as $2|\mathrm{log}(\delta a)|/N$.  
This allows a lower bound on $N$ (given the seed noise value $\delta a$) above which spontaneous emission processes can be ignored. 
On the other hand, given the number of components $N$, this estimation allows to put a lower bound for the seed noise value. 
Therefore, it is as if a certain amount of quantum noise is intrinsically present in the system. 
External sources of noise, such as interactions with the surroundings, finite temperature, etc, can only increase this value. 
Note that this is indeed a rough estimation, since Eq. (\ref{eq:decay}) is obtained by perturbative arguments.

For adiabatic processes the state of the systems under evolution follows the instantaneous ground state.
Therefore, if  changes to the Hamiltonian are performed adiabatically, say along the arrows E1 or F,  the appearance of a transient phase during the crossing of a foreigner equilibrium phase is to be expected.
In contrast, for the sudden quenches we address here, the parameters of the Hamiltonian are instantaneously switched from the start to the end point of the arrows in Fig. \ref{Pdiagram}, never taking values belonging to the equilibrium phase along the way.  
These facts render even more surprising the appearance of the long-lived transients observed here for which the  quenched dynamics qualitatively resembles the adiabatic one.
Note however that, whereas the adiabatic state of the system is fully characterized by its wave function, the dynamical equations  Eqs.~(\ref{Heff},  \ref{bmotion} ,\ref{lambda_t}) have a memory kernel. Thus an instantaneous state of the system during its evolution is characterized by its wave function as well as by the previous time dependence of the order parameter $\av{a(t)}$.

\subsection{Dynamical phase transition}

Another remarkable feature of the post-quench dynamics is the dynamical phase transition, arising for  quenches deep in the lBEC phase.     
 
As described in Sec. \ref{tolBEC}, a critical value of the quench amplitude $\Delta \varepsilon_0$ could be found that separates a fully relaxed long-time state from a regime where the system  remains with an excited state. 
The latter case is characterized by a persistent phase rotation of $\av{a}$, whereas its modulus approaches a constant value,  and is observed for a quench amplitude larger than the threshold amplitude. 
We did not observe any damping for the timescales available in our calculations, including for quenches very close to $\Delta \varepsilon_{0}^{crit}.$
The presence of a well-defined critical quench amplitude $\Delta \varepsilon_{0}^{crit}$ where the behavior of the system changes qualitatively 
is a strong evidence of a dynamical phase transition. 
Mathematically it corresponds to a non-analytical dependence of the final asymptotic state on the quench strength.

\begin{figure}[t]
\begin{centering}\center{\includegraphics[width=0.8\linewidth]{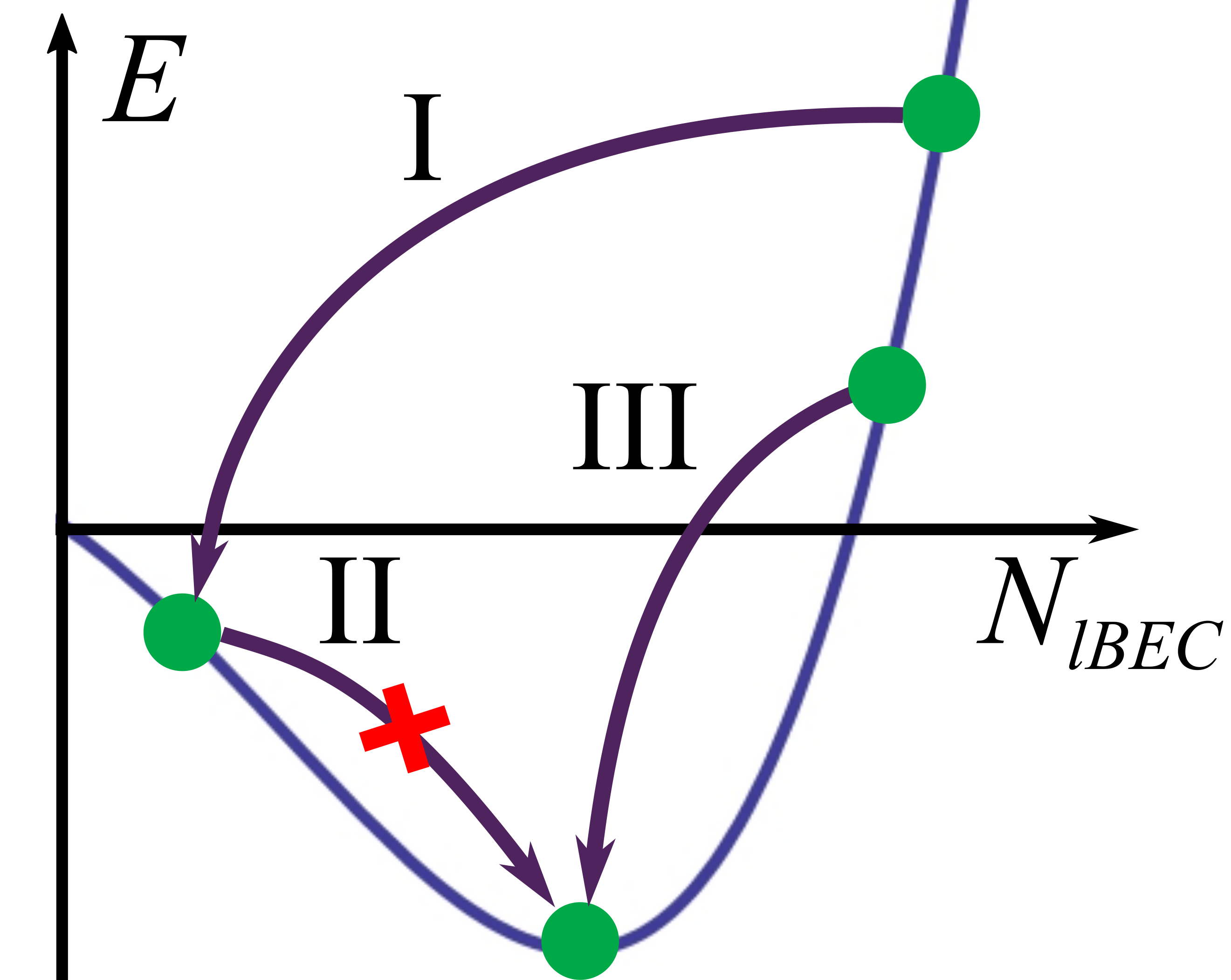}}
\par\end{centering}
\protect\caption{The illustration of how the persistent oscillating phase appear. 
The transition I corresponds to the quench within the lBEC phase with large quench amplitude (D series in the Fig. \ref{Pdiagram}). During the non-adiabatic evolution I, the impurity loses more particles, than it is required to reach the ground state. Once the state with the minimal energy is ``overshot", it cannot be reached
because emitted particles never return to the impurity (that is, the transition II is forbidden).
Transition III corresponds to the D series quench with an amplitude below the critical value. In this case the evolution is more adiabatic-like, and the system relaxes to the ground state.}
\label{potent}
\end{figure}

To explain the nature of the frictionless phase rotation, let us consider 
the possible relaxation mechanisms.
According to Eq. (\ref{bmotion}), energy and, therefore, particles transfer from the impurity to the lattice mode $k$  whenever $\av{a(t)}$ contains a non-zero Fourier component at frequency $\varepsilon_k$. 
Assuming, for the sake of the argument, that $\av{a}=a_0 e^{-i \varepsilon_k t}$, the solution of  Eq. (\ref{bmotion}) reads $\beta_k(t)=\left[ V_k \sqrt{N} a_0 t+\beta_k(0)\right] e^{-i \varepsilon_k t}$, corresponding to a linear-in-time increase of the amplitude of the $k$ mode.
This argument is only valid for short times before non-linear effects ensue. Nonetheless it provides a physical picture for the energy transfer between the impurity and the lattice with depletion of the 
Fourier component $\omega = \varepsilon_k$ in the oscillatory behavior of  $\av{a(t)}$. 
Since lattice oscillators form a continuum spectrum for $\omega>0$, the only possibility for a frictionless dynamics 
is the existence of some spectral weight of $\av{a(t)}$ in the negative frequencies. 
In this case, the order parameter effectively decouples from the continuum of single-particle lattice excitations. 
This simple picture is confirmed by our observations. 
Fig. \ref{MISFquench} shows that the order parameter 
 $\av{a}$ rotates clockwise on the complex plane at the initial stages of evolution, 
and afterwards the rotation suddenly changes to counter-clockwise direction. In this case most of the spectral weight belongs to the negative frequencies.
A small part contributing positive frequencies reveals itself in the amplitude oscillations 
of the order parameter. The latter are dissipative, so that finally a pure phase rotation  $\av{a}=a_0 e^{i \omega t}$  with $\omega>0$ is observed. The same counterclockwise rotation occurs after the quenches within lBEC phase as it can be seen from Fig. \ref{deltaeps} and from the Husimi function plot of the same data, presented in the Fig. \ref{Husimi}.

The negative part of the spectrum appears to correspond to states with less particles then the equilibrium one.
A simple explanation can be gives as follows. Consider a single impurity Hamiltonian (\ref{Himp}) with some $\varepsilon_0<0, U>0$. It has a non-monotonous spectrum: the energy $E_n$ 
decreases with the number of particles $n$ until $n$ reaches the equilibrium value
$n_0\approx-\frac{\varepsilon_0}{U}-\frac{1}{2}$. The counterclockwise phase rotation 
is  realized for a superposition state  $\ket{\Psi}=\alpha e^{-iE_{n}t}\ket{n} + e^{-iE_{n+1}t}\beta \ket{n+1}$ if $E_n<E_{n+1}$, that corresponds to $n, n+1$ being less then the equilibrium number of particles $n_0$.

The asymptotic long-time dynamics of the impurity model (\ref{generalizedH}) is determined by the number 
of bosons $N_{\infty}$ contributing the lBEC at $t\to\infty$. In $N_{\infty}$ we count the particles on the impurity itself, as well
as in the lBEC cloud located near the impurity and interacting with it. However, $N_{\infty}$ does not include all the particles of the system. As we discussed above, the relaxation processes take place at the initial stages of evolution, and the relaxation is associated with the emission of particles from the lBEC cloud. Whose particles run away from the impurity and cannot be absorbed back to the cloud. A partial runaway of bosons is obvious for the quench  between the lMI phases with $\av{n}=2$ and $\av{n}=1$ (remind that the lattice modes are not polarized in the lMI phase, so that $N_\infty$ has contributions only from bosons on the impurity). We conclude that the phase-rotating states obey a smaller $N_\infty$ then the equilibrium one.
As the collective lBEC state contains a large amount of bosons, a gradual change of $N_\infty$ 
gives rise to a continuous spectrum of the rotation frequency: smaller number of particles corresponds to faster rotation.

While the total number of particles in the system is conserved, the value of $N_\infty$ particularly depends on the emission dynamics. A dynamical phase transition separating the evolution towards the new equilibrium and the permanent phase rotating state provides an evidence that emission of particles during the after-quench evolution is essentially non-adiabatic.  After the quench with a large amplitude, the system emits more particles than it is required to reach the new equilibrium state. Since the emitted particles cannot be absorbed back to the lBEC, such an ``overshooting'' results in persistent phase rotation, as Fig. \ref{potent} illustrates. 
The smaller the quench amplitude is, the larger number of particles remains in the system and the faster phase oscillations are. Below the critical quench amplitude, the ``overshooting'' is absent.

The existence of a persistent rotating phase can also be seen as a symmetry broken ground state of an Hamiltonian in a convenient rotating frame. 
The change of frame corresponds to the gauge transformation $\tilde{a}=a e^{i\omega t}, \tilde{b}_k=b_k e^{i\omega t}$, so that $\av{\tilde{a}}$
becomes time independent for asymptoticly long times. 
The gauge transformation introduces an effective chemical potential:
\begin{equation}\label{rotH}
\begin{array}{l}
H^{\omega}=\displaystyle\sum\limits_j (\varepsilon_0+\omega) a_j^\dagger a_j + \frac{U}{2} \sum\limits_j a_j^\dagger  a_j^\dagger a_j a_j \\[15pt]
-  \displaystyle\sum\limits_{j,k} \frac{V_k}{\sqrt{N}} (a_j^{\dagger} b_k + h.c.) + \sum\limits_k (\epsilon_k+\omega) b_k^{\dagger} b_k.
\end{array}
\end{equation} 
For this Hamiltonian, the existence of a symmetry broken ground state with non vanishing $\av{\tilde{a}}$ corresponds to a persistent phase rotation in the original frame. 
These considerations imply that a finite gap in the bath spectrum $\varepsilon_k$, as well as a shift in $\varepsilon_0$, would only shift the value of $\omega$ for which the symmetry broken phase starts to appear. 
Therefore the undamped oscillatory phase is expected generically for quenches within  lBEC phase.

An analysis of the Hamiltonian (\ref{rotH}) also gives more insight about the nature on the dynamic transition. The remarkable consequence of the energy gap appeared in (\ref{rotH}) is that expression for the number of particles in the lattice cloud $\sum_k \av {b_k^\dag b_k}=N\sum_k V_k^2 \frac{|\av{a}|^2}{(\epsilon_k+\omega)^2}$ is not divergent at small $k$ for $\omega>0$, in contrast to the case of the gapless reservoir in (\ref{generalizedH}). Therefore the number of particles in the persistent rotating phase is not just quantitatively smaller than for lBEC; the difference is qualitative. We also conclude that the observed scenario of the dynamic transition cannot be realized (or at least requires a major modification) for higher dimensions $d>4$, since in that case the sum converges for a gapless reservoir as well.

Let us discuss the nature of the lBEC phase, in view of these findings. 
Strictly speaking, calling this phase a superfluid it is not well justified, as there is no evidence of superfluid dynamics in the Bose-Anderson impurity model. 
Even the persistent rotating phase also does not imply a flow of particles. 
Here,  a continuous spectrum of frictionless excitations can be seen as a reminiscent of the superfluidity in a 0D system.

Now we address the consequences of our findings for the existence of a  frictionless phase rotation regime in the finite $N$ case. We concluded that for $N \to \infty$ the dynamical transition separates the lBEC phase and persistent rotating phase, respectively containing an infinite and a finite number of particles in the cloud. An infinite number of particles in the lBEC mode was also found for the single-component Bose-Anderson impurity model \citep{Lee2010}. Moreover, a narrow peak at negative frequencies has  been observed in the equilibrium local spectral function of the zero-temperature single-component Bose-Anderson impurity model \citep{Lee2010}. Both $N=1$ and $N\to\infty$ cases share this feature, with a major part of the spectral weight belonging to the positive semi-axis. We interpret this peak as an evidence that $N=1$ model has the excitations analogous to the rotating phase state. Of course, the presented argumentation does not prove that the dynamic transition we found for $N \to \infty$ also shows up for a finite $N$. There can be at least two other possibilities. First, it might happen that a critical quench value vanishes for a finite $N$, so that the system can never relax to the new equilibrium state. In this case, no dynamic transition will be realized. Second, it is not clear if the symmetry-broken state is stable for a finite $\omega$, or the rotating state will decay to $\av{a}=0$ because of quantum fluctuations in a finite $N$ system. In the latter case the dynamic transition will take the place, but its properties will be different from the $N=\infty$ system. Solving these questions is will be a subject of the future work.

Finally let us address the differences and similarities of the persistent rotation phase found here with that found in the dynamics of the superconducting order parameter in the BCS model \cite{Yuzbashyan2006}. Depending  on the initial quench, the BCS system can also show a relaxation to the new steady state or persistent oscillations of the order parameter. 
There are however remarkable differences from the case studied here: 
(i) In the B-SIAM interactions and quenches are local, therefore the energy injected to the system is not extensive and, as a consequence, the relaxation to the new equilibrium ground state is to be expected. On the contrary, in the BCS case the system never relaxes to the ground-state but rather to a finite energy state. 
(ii) The analysis of the BCS model \cite{Yuzbashyan2006} is based on its mapping to a classical dynamical system within Anderson pseudospin representation. 
Furthermore integrability is exploited. In our case the impurity part of the Hamiltonian is treated exactly and the integrals of motion can hardly be established. 
(iii) As a result of the mean-field treatment in the BCS case the system is effectively Gaussian. In our case local interactions are treated exactly and are highly non Gaussian. 
(iv) Finally, as discussed above, our system is robust against $1/N$ corrections whereas in the BCS phase the possible effect of anharmonic corrections terms is not clear.

\section{Conclusions}
\label{sec:5}

We have studied quenches in the $N$-component Bose-Anderson model, which in the $N \to \infty$ limit allows a numerically exact solution. 
Such treatment is possible because the $N \to \infty$ limit corresponds to the suppression of quantum fluctuation of the bath degrees of freedom rendering exact a mean-field-like treatment of the bath modes. 
As local interaction effects are kept at the impurity level, our approach is able to capture a set of interesting phenomena arising due to quantum many-body correlations. 
The model exhibits a rich set of dynamical regimes at different timescales as a result of the interplay between local and bath degrees of freedom:
short times are related with the internal impurity degrees of freedom, whereas the long time dynamics is determined by the collective modes. 

The equilibrium zero-temperature phase diagram of the B-SIAM, consists of a set of lMI lobes and a lBEC phase. 
This non-trivial phase diagram allows for lMI-lMI or lBEC-lBEC quenches that cross the complementary phase. These quenches are observed to give rise to long-lived transient phases. 
The transient lBEC-like phase, formed while quenching between the lMI states, is explained using analytic consideration valid for a weak bath-impurity coupling. 
A parallel is drawn with the superradiant regime arising in the celebrated Dicke model \cite{Dicke}, which describes a large number of two-level systems interacting with a bosonic field.
Long-lived lMI-like plateaus are found for lBEC-to-lBEC quenches that cross Mott lobes. 

A non-decaying mode is found throughout lBEC phase. 
This negative frequency excitation is well separated from the spectral continuum located at positive energies. 
The existence of such long-lived mode ensures the possibility of observing persistent phase rotations of the order parameter for quenches with a large enough amplitude.
As its characteristic frequency is negative, the phase of the order parameter rotates counterclockwise, whereas all clockwise rotating, i.e. positive energy, modes are damped due to the interaction with the bath degrees of freedom.
The non-decaying mode is unique and corresponds to a well defined frequency at each point of the phase space of the model. 
The particular value of the frequency depends on the total particle number.  This fact, together with the conservation of particle number throughout the evolution that is fixed by the initial state, explains that quenches to the same point in phase space give rise to persistent oscillations with different frequencies.

A dynamical phase transition is found, as a function of the quench amplitude, separating the regimes where the evolution either attains the equilibrium lBEC phase or retains a persistent oscillating phase. 
The high accuracy of the calculations has allowed us to obtain strong numerical evidence that the two regimes are separated by a singular line, so that 
a true dynamical transition - rather them a crossover - occurs.

Our findings, regarding the existence of a persistent phase rotation mode, are robust to $1/N$ corrections and should be present down to $N=1$ since a resonance at negative energies has already been reported in the spectral function of the B-SIAM \cite{Lee2010}. This phenomenon is generic and should be observed ubiquitously in open quantum systems whenever conservation laws permit the existence of isolated spectral modes. 

More general implication of the work will require a consideration of the several impurities connected to the different lattice sites. The main difference of these systems from that considered in our work is that 
their finite size makes the persistent (superfluid) current possible. Thus, they can be considered as (quantum) memory cells  and their relaxation dynamics becomes of potential technological importance.

\section{Acknowledgements}
We acknowledge useful discussions with P.I. Arseev, I.S. Krivenko, Georg Rohringer and V.I. Yudson. 
P.R. acknowledges financial support from FCT thought the contract Ref. IF/00347/2014/CP1214/CT0002 under the IF2014 program.
D.V.C., Y.E.S., and  A.N.R. thank the Dynasty foundation and RFBR (grants 14-02-01219 and 16-32-00554)  for financial support. Modeling of the quenches to lMI phase (section II B) and the analysis of the transient phases
(section IV A) was funded by the RSF grant 16-42-01057.


\begin{appendix}

\section{Dynamics in the large $N$ limit} 
\label{sec:LN}

\subsection{Diagramatics}

Let us construct a formal $1/N$ expansion. The lattice degrees of freedom can be integrated out of the initial Hamiltonian (\ref{Himp}), yelding the action
\begin{equation}
S=\sum S_{SI} [a^\dag_j a_j]+\sum_j a_{j t}^\dag \Delta_{t-t'} \sum_{j'} a_{j't'},
\end{equation} 
where $S_{SI}$ is the single-impurity action corresponding to the Hamiltonian (\ref{Himp}), and $\Delta_t=\frac{V_k}{\sqrt{N}} (i\partial_t -\varepsilon_k)^{-1} \frac{V_k}{\sqrt{N}}$ is the hybridization function reflecting the hopping and lattice dispersion properties; the integration over time-arguments is assumed.
According to the equation (\ref{lambda_t}), $\lambda_t=\int \Delta_{t-t'} \av{a_{t'}}$, and therefore the action can be re-written as
\begin{equation}
S=\sum S_{SI}^{eff} [a^\dag_j a_j]+\sum_j \tilde{a}_{j t}^\dag \Delta_{t-t'} \sum_{j'} \tilde{a}_{j't'},
\end{equation} 
where $S_{SI}^{eff}$ corresponds to the effective impurity Hamiltonian (\ref{Heff}), and $\tilde{a}_j=a_j-\av{a_j}$ describes the displacement of the impurity $j$ from the mean-value. Neglecting the hybridization term in this formula, one restores the mean-field theory presented above. The deviations from this result can be estimated considering the serial expansion in $\Delta$. Let us formally integrate out all the impurities, except a single one labeled $j_0$. 
In the zeroth order, the impurities are uncorrelated, and the result of such an integration is just 
\begin{equation}
S^{(0)}_{SI}[a_{j_0}^\dag, a_{j_0}]=S_{SI}^{eff}[a_{j_0}^\dag, a_{j_0}].
\end{equation}  
The first-order is contributed only by the diagonal term
\begin{equation}
S^{(1)}_{SI}[a_{j_0}^\dag, a_{j_0}]=\tilde{a}_{j_0 t}^\dag \Delta_{t-t'} \tilde{a}_{j_0 t'} 
\end{equation}
The value of $\Delta$ scales as $1/N$, so the first-order correction 
is also proportional to $1/N$. The 
effect of mutial correlations between the different impurities arises in the second order. 
The second-order correction to the effective impurity action obeys the form
\begin{equation}
S^{(2)}_{SI}[a_j^\dag, a_j]=\tilde{a}_{j t}^\dag \left(\sum_{j'} \Delta_{t-t_1} \av{\tilde{a}_{j' t_1} \tilde{a}_{j' t_2}^{\dag}} \Delta_{t_2- t'} \right) \tilde{a}_{j t'}.
\label{S2}
\end{equation}
Since the averages $\av{\tilde{a}_{j'} \tilde{a}_{j''}^{\dag}}$ for $j'\neq j''$ are vanished in the zeroth-order, they carry an additional smallness and should not be accounted in the right-hand side of (\ref{S2}). The sum over $j'$ contains $N$ equal terms, so the entire expression scales as $1/N$. It is easy to see that higher expansion terms also carry $1/N$ prefactor. This concludes the proof that the mean-field theory is exact in the limit of infinite number of impurities.

\subsection{Path integral} 
In this section we derive the $1/N$ expansion and show that Eqs.~(\ref{Heff},\ref{lambda_t}) are exact in the large $N$ limit.
We start by writing the generating function for the Hamiltonian (\ref{generalizedH}) on the Keldysh contour $\gamma$: 
\begin{eqnarray}
Z\left[\xi\right] & = & \int\prod_{j}Da_{j}\prod_{k}Db_{k}\,e^{-i\left\{ \sum_{j}S_{j}\left[\xi\right]+S_{\text{bath}}+S_{\text{int}}\right\} }
\end{eqnarray}
where $\xi$'s are source fields and  
\begin{eqnarray}
S_{\text{loc},j}\left[\xi\right] & = & - \int dz dz' \sum_{j} a_{j}(z)^{\dagger}G_{0}^{-1}(z,z')a_{j}(z') \nonumber \\
 & &+\frac{1}{2}\int_{\gamma}dz\,\sum_{j} U\left(z\right)a_{j}^{\dagger}(z)a_{j}^{\dagger}(z)a_{j}(z)a_{j}(z) \nonumber\\
 & & - \int_{\gamma}dz \sum_{j} \left[ \xi^{\dagger}(z) a_{j}(z)+a_{j}^{\dagger}(z)\xi(z) \right]\\
S_{\text{bath}} & = & -\int_\gamma dz \sum_{k}b_{k}^{\dagger}(z) g_{k}^{-1}(z,z') b_{k}(z')\\
S_{\text{int}} & = & -\int_{\gamma}dz\,\sum_{k,j} \frac{V_{k}}{\sqrt{N}}\left[b_{k}^{\dagger}\left(z\right)a_{j}\left(z\right)+a_{j}^{\dagger}\left(z\right)b_{k}\left(z\right)\right]
\nonumber \\ 
\end{eqnarray}
are respectively the impurity, bath and interaction terms in the action. $G_{0}$ and $g_{k}$ are the propagators of the impurity 
and of the bath in the absence of coupling or interactions. 
Integrating out the bosonic bath one obtains an action solely in terms of the $a$ fields
\begin{eqnarray}
Z\left[\xi\right] & = & \int Da\,e^{-i\left\{ \sum_{j}S_{\text{loc},j}\left[\xi\right]+\frac{1}{N}\sum_{jj'}a_{j}^{\dagger}\Xi^{-1}a_{j'}\right\} }
\end{eqnarray}
where
\begin{eqnarray}
\Xi^{-1}\left(z,z'\right) & = & \sum_{k}V_{k}\left(z\right)g_{k}\left(z,z'\right)\bar{V}_{k}\left(z'\right)
\end{eqnarray}
with $\Xi^{\dagger}=\Xi$. 
We now introduce an Hubbard-Stratonovich field $\lambda$ to decouple the term in $\Xi$ which factorizes the 
impurity action into independent components. 
Since the action is symmetric with respect to component exchange the total action is $N$ times the one of a single component.   
Therefore, we obtain  
\begin{eqnarray} \label{partition_N}
Z\left[\xi\right] & = & \int D\lambda\,e^{-i N S\left[\lambda\right]}
\end{eqnarray}
where
\begin{eqnarray} \label{action_N}
S\left[\lambda\right] & = &  -\int_\gamma dzdz' \lambda^{\dagger}(z)\Xi(z,z')\lambda(z')+F\left[\lambda\right] 
\\
\label{free_energy_N}
F\left[\xi+\lambda\right] & = & i\,\ln\int Da\,e^{-iS_{\text{loc}}\left[\xi+\lambda\right]}.
\end{eqnarray}
Note that the fact that the action, in Eq.~(\ref{partition_N}), is multiplied by $N$  makes the integral amenable to be treated by a saddle-point approximation 
whenever $N$ is large. 

We now proceed to derive the saddle-point conditions setting the sources $\xi$ to zero. Varying the action in order to $\lambda^\dagger$ we obtain
\begin{eqnarray}
\delta_{\lambda^{\dagger}(z)} S\left[\lambda\right] & = & -\int_\gamma dz' \Xi (z,z') \lambda(z')+\delta_{\lambda^{\dagger}(z)}F\left[\lambda\right]. \nonumber \\
& &
\end{eqnarray}
From the definition in Eq.(\ref{free_energy_N}) we can identify 
$\delta_{\lambda^{\dagger}(z)}F\left[\lambda\right] = - \av{a\left(z\right)}$, 
where the mean value $\av{...}$ is taken with respect to the Hamiltonian (\ref{Heff}). 
Setting the variation of the action to zero and inverting the kernel $\Xi$ thus we obtain:
\begin{eqnarray}
\lambda\left(z\right) & = & -\int_{\gamma}dz'\,\Xi^{-1}\left(z,z'\right).\av{a\left(z'\right)}.
\end{eqnarray}
In order to obtain the equations in the main text we have to evaluate the former expression in the real time branch of the Keldysh contour:
\begin{eqnarray} \label{lambda_t_0}
\lambda\left(t\right)&=& i \int_{0}^{\beta}d\tau\Xi^{-1\urcorner}\left(t,\tau\right) \av{a\left(\tau\right)} \\
 & &-\int_{0}^{t}dt'\left[\Xi^{-1>}\left(t,t'\right)-\Xi^{-1<}\left(t,t'\right)\right]\av{a\left(t'\right)} \nonumber 
\end{eqnarray}
Identifying the different components of $\Xi$, assuming the bath is at thermal equilibrium at $t=0$ with inverse temperature $\beta$, we get
\begin{eqnarray}
\Xi^{-1M}(\tau,\tau')&=&-i \sum_k V_{k}\left(0\right)\bar V_{k}\left(0\right) \left[n_{b}(\epsilon_k)+\Theta\left(\tau-\tau'\right)\right]e^{-\epsilon_k \left(\tau-\tau'\right)}
\nonumber\\
\\
\Xi^{-1\urcorner}\left(t,\tau'\right)&=&-i \sum_k  V_{k}\left(t\right)\bar V_{k}\left(0\right) n_{b}(\epsilon_k)e^{-i\epsilon_k\left(t+i\tau\right)}\\
\Xi^{-1>}\left(t,t'\right)&=&-i \sum_k  V_{k}\left(t\right)\bar V_{k}\left(t'\right) \left[n_{b}(\epsilon_k)+1\right]e^{-i\epsilon_k\left(t-t'\right)}\\
\Xi^{-1<}\left(t,t'\right)&=&-i \sum_k  V_{k}\left(t\right)\bar V_{k}\left(t'\right) n_{b}(\epsilon_k)e^{-i\epsilon_k\left(t-t'\right)}
\end{eqnarray}
with $n_b$ the Bose-Einstein distribution. With this assumptions, and using that along the thermal branch $\av{a\left(\tau\right)}=\av{a\left(0\right)}$ , Eq.(\ref{lambda_t_0}) becomes
\begin{eqnarray} \label{lambda_t_1}
\lambda\left(t\right)&=&\sum_{k}V_{k}\left(t\right)\frac{e^{-it\epsilon_{k}}}{\epsilon_{k}}\bar V_{k}\left(0\right)\av{a\left(0\right)} \\
& &+i\sum_{k}\int_{0}^{t}dt'V_{k}\left(t\right)e^{-i\left(t-t'\right)\epsilon _{k}}\bar V_{k}\left(t'\right)\av{a\left(t'\right)} \nonumber
\end{eqnarray}
which becomes Eq.(\ref{lambda_t}) when $V_{k}$ does not depend on time. In the limit $N\to\infty$ the saddle-point value of $\lambda$ obtained 
by Eq.(\ref{lambda_t_1}) becomes exact.

\end{appendix}

\bibliography{biblio}

\end{document}